\documentclass{jpsj3}
\usepackage{txfonts}

\title{Effective field theory for two-species bosons in an optical lattice:  \\
Multiple order, the Nambu-Goldstone bosons,
the Higgs mode and vortex lattice}

\author{Yoshihito Kuno, Keita Suzuki, and Ikuo Ichinose}
\inst{Department of Applied Physics, Nagoya Institute of Technology,
Nagoya, 466-8555 Japan} 

\abst{In the previous papers, we studied the bosonic t-J mode and derived an effective 
field theory, which is a kind of quantum XY model.
The bosonic t-J model is expected to be realized by experiments 
of two-component cold atoms in an optical lattice.
In this paper, we consider a similar XY model that describes phase diagram
of the t-J model with a mass difference.
Phase diagram and critical behavior of the quantum XY model are clarified 
by means of the Monte-Carlo simulations.
Effective field theory that describes the phase structure and low-energy
excitations of the quantum XY model is derived.
Nambu-Goldstone bosons and the Higgs mode are studied by using
the effective field theory and interesting findings are obtained for the
system with multiple order, i.e., Bose-Einstein condensations and pseudo-spin
symmetry.
We also investigate physical properties of the quantum XY model 
in an effective magnetic field that is realized by rotating the optical lattice, etc.
We show that low-energy states of the system strongly depend on the
strength of the ``magnetic field".
For some specific strength of the magnetic field, vortex lattice forms
and the correlation function of the bosons exhibits solid like behavior,
which is a kind of Bose-Einstein condensation.}

\kword{bosonic t-J model, cold atom, optical lattice, supersolid,
phase separation}

\begin{document}
\maketitle

\section{Introduction}

Recently cold atomic systems are one of the most actively studied fields
in physics\cite{optical}.
The versatility of cold atom systems offers new methods for investigating 
problems that are difficult to be studied by means of conventional methods.
In particular, the cold atomic system in an optical lattice is sometimes regarded
as a ``quantum simulator" and it is expected to give important insights into
properties of strongly-correlated many-body systems\cite{lw}.
The cold atomic systems in an optical lattice are highly controllable, e.g., the
dimension and type of lattice are controlled by the setup of the experimental 
apparatus, the interactions between atoms are freely controlled by the Feshbach
resonance, etc.

It is now widely accepted that a single-species boson system in an optical
lattice is described by the Bose-Hubbard model\cite{BHM}.
The Mott-superfluid phase transition, which was observed in the experiments\cite{MS},
is well described by the Bose-Hubbard model.
Multi-species (multi-component) boson systems are expected to 
have a rich phase structure and are realized by, e.g., $^{85}$RB -$^{87}$Rb,
$^{87}$RB -$^{41}$K mixture\cite{Rb2,RbK}.
These multi-component systems were theoretically studied by various methods.
The two-component Bose-Hubbard model at commensurate fillings
has been studied in e.g., Refs.\cite{altman,sansone,Hof}
by the mean-field-theory (MFT) type approximations and the numerical methods.
It was predicted that interesting states including the super-counter-fluid,
supersolid (SS), etc, form in certain parameter regions.
Doped two-component hard-core Bose-Hubbard model was studied by
using the Monte-Carlo (MC) simulations\cite{moreno}, and it was shown
that five distinct phases can exist.

In this paper,
we are interested in the strong replusive case of the two-component 
model, which is a bosonic counterpart of the strongly-correlated electron systems 
like the high-$T_c$ materials\cite{LNW}.
It is expected that various phases appear in that system at incommensurate
particle density.
The results obtained for that system may give important insight into the 
phase diagram of the fermionic counterpart.
In the previous papers\cite{BtJ,BtJ1}, we showed that the strong-repulsive
Bose-Hubbard
model is well described by the bosonic t-J model\cite{BtJ2} and studied its
phase diagram, etc.
To this end, we employed the path-integral formalism with the slave-particle
representation.
In Ref.\cite{BtJ}, we studied the finite-temperature properties of the
bosonic t-J model on a stacked triangular lattice.
In particular we were interested in the anti-ferromagnetic $J$-couplings
that generates the frustration.
By means of the MC simulations, the phase diagrams of the system were
investigated rather in detail.
However to perform the MC simulations, we ignored the Berry phase in the
action assuming that the existence of the Berry phase does not influence
the {\em finite-temperature} phase diagram substantially.
On the other hand in Ref.\cite{BtJ2}, we studied the ground-state 
phase diagram of the bosonic t-J model on a square lattice.
We first integrated out the amplitude degrees of freedom of the slave particles
in order to make the action of the model positive-definite.
The resultant model describes the phase degrees of freedom of each 
atom and hole and we call it ``quantum XY model".
As the action of the quantum XY model (qXY model) is positive-definite,
a straightforward application of the Monte-Carlo (MC) simulation to it is
possible.
Furthermore, a low-energy effective field theory was obtained by means
of a ``Hubbard-Stratonovich" transformation. 
Phase diagram of the qXY model and low-energy excitations, e.g., Nambu-Goldstone
bosons, were studied analytically by using the effective field theory.
 
In this paper, we shall extend the previous studies\cite{BtJ,BtJ1}.
The extension is three-fold.
\begin{enumerate}
\item a finite mass difference of the $a$ and $b$-atoms
\item a finite $J_z$-term in the qXY model and its effect on supersolid
\item effects of an external (synthetic) magnetic field
\end{enumerate}

This paper is organized as follows.
In Sec.II, we introduce the bosonic t-J model and the qXY model.
Relation between the bosonic t-J model and the Bose-Hubbard-J model
is also explained.
Phase diagram of the qXY model with a mass difference is
obtained by the MC simulations.
Topological excitations, i.e., vortices are also studied numerically.
Section III is devoted for study of the supersolid that forms 
as a result of sufficiently large $J_z$-term of the pseudo-spin interactions.
Parameter region of the SS in the phase diagram is clarified by the numerical study.
In Sec.IV, we derive an low-energy effective field theory taking account of
the $J_z$-term.
The obtained phase diagram of the qXY model by the numerical study in Sec.II is
re-derived by using the effective potential of the effective theory.
We also study the low-energy excitations including the Nambu-Goldstone boson 
and the Higgs mode, and obtain interesting results.
In Sec.V, we study effects of the synthetic magnetic field to
the superfluid (SF) phase.
We show that the SF is destroyed by a small amount of the magnetic field.
However, we also find that there exist stable SFs at some specific strength of 
the magnetic field.
Various correlation functions exhibit unusual behaviors there.
Detailed study on these states is given and it is found that some specific
vortex lattices form there.
Section V	 is devoted for conclusion.

\section{Phase diagram of quantum XY model for $\mbox{t-J}$ model with mass
difference}
\setcounter{equation}{0}

In this section, we shall study the phase diagram of the bosonic t-J model
with a mass difference.
Hamiltonian of the system is given as 
\begin{eqnarray}
H_{\rm tJ}&=&-\sum_{\langle i,j\rangle} (t_a a^\dagger_{i}a_j
+t_b b^\dagger_{i}b_j+\mbox{h.c.})
+J_z\sum_{\langle i,j\rangle}S^z_{i}S^z_j  
 -J\sum_{\langle i,j\rangle}(S^x_{i}S^x_j+S^y_{i}S^y_j),
\label{HtJ}
\end{eqnarray}
where $a^\dagger_i$ and $b^\dagger_i$ are 
boson creation operators at site $i$ of a square lattice.
Pseudo-spin operator $\vec{S}_i$ is given as  
$\vec{S}_i={1 \over 2}B^\dagger_i\vec{\sigma}B_i$ with
$B_i=(a_i,b_i)^t$, and $\vec{\sigma}$ is the Pauli spin matrix.
In the original t-J model, the doubly-occupied state is excluded at each site.
In the present study, we extend the above constraint to the one such that
the total number of $a$-atom and $b$-atom at each site is less than the (freely) 
assigned value $N$.
Furthermore we add the following potential term $H_V$ that controls fluctuations
of the particle numbers at each site,
\begin{equation}
H_V={V_0 \over 4}\sum_i\Big((a_i^\dagger a_i-\rho_{ai})^2+
(b_i^\dagger b_i-\rho_{bi})^2\Big),
\label{HV}
\end{equation}
where $\rho_{ai}+\rho_{bi}\leq N$ and $V_0$ is a positive parameter.
Parameter $V_0$ is obviously related to the on-site repulsion of atoms,
but we regard it as a free parameter with the energy dimension.
The Heisenberg ``pseudo-spin" terms in $H_{\rm tJ}$ (\ref{HtJ})
represent the interactions between the $a$ and $b$-atoms
at nearest-neighbor (NN) sites.
Experimental realization of the above ``nonlocal" interactions between atoms 
is interesting and important.
The $J_z$-term obviously gives an inter and intra-species interactions
of the $a$ and $b$-atoms at NN sites.
Besides the ordinary Feshbach resonance, dipolar interaction might be
useful for realizing the NN interaction\cite{dipolar}.
For $J_z>0$ and in a bipartite lattice, the checkerboard (CB) configuration with
a Ising type order is enhanced, whereas for $J_z<0$, a homogeneous 
configuration is favored.
On the other hand, the $J$-term controls the relative phase of the $a$
and $b$-atoms' condensates.
For example the FM (AF) interaction $J>0$ $(J<0)$ prefers 
$\langle a^\dagger_i \rangle=\langle b^\dagger_i \rangle$ 
($\langle a^\dagger_i \rangle=-\langle b^\dagger_i \rangle$).
The $J$-term 
$(S^x_{i}S^x_{i+\hat{k}}+S^y_{i}S^y_{i+\hat{k}}) 
\propto (a^\dagger_ib_ib^\dagger_{i+\hat{k}}a_{i+\hat{k}}+\mbox{H.c.})$ \
($\hat{k}$=unit vector)
is realized experimentally by putting auxiliary fermions
(or bosons) on the link $(i,i+\hat{k})$ of the optical lattice\cite{Jterm}.
These fermions are coupled with the bosons $a_i$, $b_i$, etc via scattering terms as 
\begin{eqnarray}
\sum_{j=i,i+\hat{k}}a^\dagger_jb_j\phi^\dagger_{i,\hat{k}}\varphi_{i,\hat{k}}
+\mbox{H.c.},
\label{link}
\end{eqnarray}
where $\phi_{i,\hat{k}}$ and $\varphi_{i,\hat{k}}$ are annihilation operators
of distinct internal states of the fermion (or boson).
By making a large energy difference between the two states 
$\phi^\dagger_{i,\hat{k}}|0\rangle$ and 
$\varphi^\dagger_{i,\hat{k}}|0\rangle$ as
\begin{equation}
H_f^0=\sum_{i,\hat{k}}\Big(m_\phi \phi^\dagger_{i,\hat{k}}\phi_{i,\hat{k}}
+m_\varphi \varphi^\dagger_{i,\hat{k}}\varphi_{i,\hat{k}}\Big), \;\;\;
m_\phi \gg m_\varphi,
\label{fermion}
\end{equation}
the interaction (\ref{link})
can be treated as a perturbation, and the second-order perturbation expansion
gives the $(S^x_{i}S^x_{i+\hat{k}}+S^y_{i}S^y_{i+\hat{k}})$-term.
The parameter $J$ in Eq.(\ref{HtJ}) is given by the the overlap integral of the
Wannier functions of each atom.
The model given by Eqs.(\ref{HtJ}) and (\ref{HV}) $H_{\rm tJ}+H_V$
without the local constraint of particle number
should be called {\em Bose-Hubbard-J model}.

Low-energy effective model for $H_{\rm tJ}+H_V$
is obtained by integration out the amplitude mode of $a^\dagger_i$ and $b^\dagger_i$
in the path-integral formalism by using a {\em slave-particle representation} 
and the qXY model for the phase degrees of freedom is derived.
In the previous papers\cite{BtJ,BtJ1}, we considered the specific case $t_a=t_b$ and 
showed that the obtained qXY model well describes low-energy properties of 
the original t-J model.
In this section, we shall continue the previous study and consider the case
$t_a\neq t_b$.
We clarify the phase diagram and low-energy excitations including Nambu-Goldstone
bosons and Higgs particles.
Topological excitations in each phase are also studied.

For the bosonic t-J model on the square lattice with $J>0$ and $J_z=0$,
action of the derived qXY model is given as follows,
\begin{equation}
A_{\rm Lxy}=A_{{\rm L}\tau}+A_{\rm L}(e^{i\Omega_\sigma},e^{-i\Omega_\sigma}), 
\label{AL1}
\end{equation}
where
\begin{eqnarray}
A_{{\rm L}\tau}=c_\tau\sum_{r} \sum_{\sigma=1}^3
\cos (\omega_{\sigma,r+\hat{\tau}}-\omega_{\sigma r}+\lambda_r),  
\label{Atau}
\end{eqnarray}
and 
\begin{eqnarray}
A_{\rm L}(e^{i\Omega_\sigma},e^{-i\Omega_\sigma}) =
-\sum_{\langle r,r'\rangle}
\Big(C^a_3\cos (\Omega_{2r}-\Omega_{2r'})  
 +C^b_3\cos (\Omega_{3r}-\Omega_{3r'}) 
+C_1\cos (\Omega_{1r}-\Omega_{1r'})\Big).
\label{AL2}
\end{eqnarray}
We have introduced a lattice for the imaginary time.
Then in Eqs.(\ref{Atau}) and (\ref{AL2}), $r$ denotes 
{\em site of the space-time cubic lattice},
$\hat{\tau}$ is the unit vector in the direction of the imaginary time,
and $\langle r,r'\rangle$ denotes the NN sites in the 
{\em 2D spatial lattice}.
$\lambda_r$ is the Lagrange multiplier field for the local constraint of the
particle number at each site in the t-J model.
In the homogeneous distribution $\rho_{ai}=\rho_a$ and $\rho_{bi}=\rho_b$,
the parameters are related to the original ones as 
\begin{eqnarray}
&& c_\tau={1 \over V_0\Delta\tau},   \nonumber \\
&& C_1=4J\rho_a^2\rho_b^2\Delta \tau \propto {J/(c_\tau V_0)}, \nonumber  \\
&& C^a_3={t_a \over 2}\rho_a(N-\rho_a-\rho_b)\Delta \tau  
\propto {t_a/(c_\tau V_0)},  \nonumber \\
&& C^b_3={t_b \over 2}\rho_b(N-\rho_a-\rho_b)\Delta \tau  
\propto {t_b/(c_\tau V_0)},
\end{eqnarray}
where $\Delta\tau$ is the lattice spacing of the imaginary time.
It should be remarked that $c_\tau, \cdots, C^b_3$ are dimensionless
parameters. (We have put $\hbar=1$.)
In Eq.(\ref{AL2}), the dynamical variables are 
$$
\Omega_{1 r}=\omega_{1 r}-\omega_{2 r}, \
\Omega_{2 r}=\omega_{1 r}-\omega_{3 r}, \
\Omega_{3 r}=\omega_{2 r}-\omega_{3 r},
$$
where $\omega_{\alpha r} \;(\alpha=1,2,3)$ are phases of the slave particles,
and the above variables are related with the original ones as
\begin{equation}
S^x_r+iS^y_r \propto e^{i\Omega_{1r}},  \; \;
a_r \propto e^{i\Omega_{2r}}, \; \;
b_r \propto e^{i\Omega_{3r}}. 
\label{Sab}
\end{equation}
Then the partition function $Z$ is given as follows by the path-integral formalism,
\begin{equation}
Z=\int [d\omega_{\alpha r} d\lambda_r]e^{A_{\rm Lxy}}.
\label{Z}
\end{equation}

We numerically studied the model defined by 
Eqs.(\ref{AL1}) and (\ref{Z}) with the value of $c_\tau$ fixed
by calculating the ``internal energy" $E$ and ``specific heat" $C$ as 
a function of $C_1$ and $(C^a_3, C^b_3)$,
\begin{eqnarray}
E&=&\langle A_{\rm Lxy} \rangle/L^3, \nonumber  \\
C&=&\langle (A_{\rm Lxy}-E)^2 \rangle/L^3,
\label{EC}
\end{eqnarray}
where $L$ is the linear size of the 3D cubic lattice with the periodic boundary
condition.
In order to identify various phases, we also measured the 
pseudo-spin and boson correlation functions that are given by,
\begin{eqnarray}
&& G_{\rm S}(r)={1 \over L^3}\sum_{r_0} \langle 
e^{i\Omega_{1r_0}}e^{-i\Omega_{1,r_0+r}}\rangle, \nonumber \\
&& G_{a}(r)={1 \over L^3}\sum_{r_0} \langle 
e^{i\Omega_{2r_0}}e^{-i\Omega_{2,r_0+r}}\rangle, \nonumber \\
&&G_{b}(r)={1 \over L^3}\sum_{r_0} \langle 
e^{i\Omega_{3r_0}}e^{-i\Omega_{3,r_0+r}}\rangle,
\label{CF1}
\end{eqnarray}
where sites $r_0$ and $r_0+r$ are located in the same spatial 2D lattice,
i.e., they are the equal-time correlators.
For example, if $G_a(r)\rightarrow\mbox{finite}$ as $r\rightarrow \infty$,
we judge that Bose-Einstein condensation (BEC) of the $a$-atom takes place.

For numerical simulations, 
we employ the standard Monte-Carlo Metropolis algorithm
with local update\cite{Met}.
The typical sweeps for measurement is $(30000 \sim 40000)\times (10$ samples),
and the acceptance ratio is $40\% \sim 50\%$.
Errors are estimated from 10 samples with the jackknife methods.

\begin{figure}[t]
\begin{center}
\includegraphics[width=10cm]{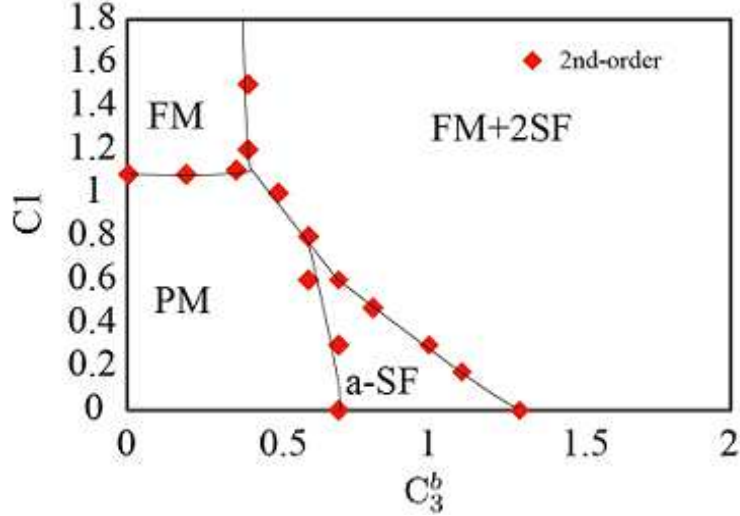}
\vspace{-0.3cm}
\caption{(Color online)
Phase diagram of the qXY model (\ref{AL1}).
There are four phases, i.e., paramagnetic (PM), ferromagnetic (FM), 
superfluid of $a$-atom ($a$-SF), and superfluid of $a$ and $b$-atoms 
accompanying the ferromagnetic order (FM+2SF).
Locations of the phase transitions are determined by the calculation of 
system size $L=16$.
The numbers in parentheses indicate the number of Nambu-Goldstone bosons
in each phase.
}
\vspace{-0.5cm}
\label{PD1}
\end{center}
\end{figure}
\begin{figure}[h]
\begin{center}
\includegraphics[width=5cm]{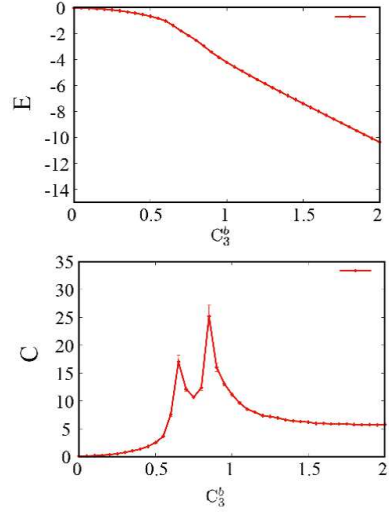}
\vspace{-0.3cm}
\caption{(Color online)
Internal energy  and specific heat as a function of $C^b_3$ for $C_1=0.3$.
Behavior of $C$ indicates that there exist two second-order phase
transitions.
$L=16$.
}
\vspace{-0.5cm}
\label{EC1}
\end{center}
\end{figure}

\begin{figure}[h]
\begin{center}
\includegraphics[width=9cm]{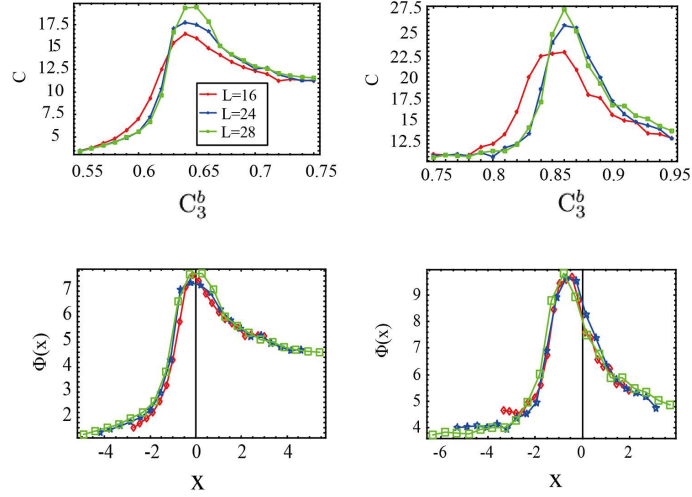}
\vspace{-0.3cm}
\caption{(Color online)
Finite size scaling for two phase transitions in Fig.\ref{EC1}.
$\Phi(x)$ is the scaling function in Eq.(\ref{FSS}).
}
\vspace{-0.5cm}
\label{fig:FSS}
\end{center}
\end{figure}

We first study a simple case in which $\rho_{ai}=\rho_{bi}=\rho$ as both the ferromagnetic
$J$-term and the hopping terms prefer the homogeneous distribution.
The case with $J_z>0$ will be studied in Sec.III.
In the practical calculation, we put $C^a_3=2C^b_3$.
We show the phase diagram obtained by the MC simulations in Fig.\ref{PD1}.
There are four phases, the paramagnetic (PM) phase without any long-range
orders (LRO), the ferromagnetic (FM) state that exhibits a FM order but 
neither $a$ nor $b$-atom Bose condenses there.
The FM state forms as a result of condensation of the bi-atom composite 
$\langle a_ib^\dagger_i\rangle \neq 0$, and is sometimes called super-counter-fluid.
There are two other phases, i.e.,
the state of the BEC of the $a$-atom without the FM order, and finally 
the FM state with BECs of both the $a$ and $b$-atoms.
Order of the phase transitions is also indicated in Fig.\ref{PD1},
and some typical behaviors of $E$ and $C$ near the phase boundary
are shown in Fig.\ref{EC1}.
We also show the result of the finite-size scaling (FSS) for two 
second-order phase transitions in Fig.\ref{EC1}.
In the FSS, the specific heat $C$ is parameterized as 
\begin{equation}
C_L(\epsilon)=L^{\sigma/\nu}\Phi(L^{1/\nu}\epsilon),
\label{FSS}
\end{equation}
where $\nu$ and $\sigma$ are critical exponents, 
$\epsilon=(C^b_3-C^b_{3\infty})/C^b_{3\infty}$ with $C^b_{3\infty}=$ 
the critical coupling for
$L\rightarrow \infty$, and $\Phi(x)$ is the scaling function.
See Fig.\ref{fig:FSS}.
For the first phase transition shown in Fig.\ref{EC1}, 
$C^b_{3\infty}=0.645, \ \nu=0.95$ and
$\sigma=0.28$, whereas for the second, $C^b_{3\infty}=0.876, \ \nu=0.88$ and
$\sigma=0.32$.

\begin{figure}[t]
\begin{center}
\includegraphics[width=8cm]{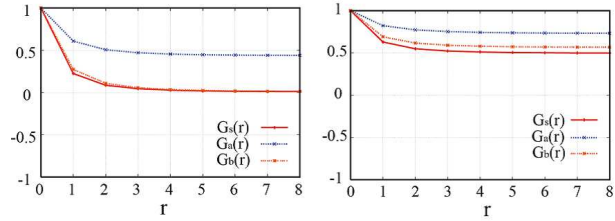}
\vspace{-0.3cm}
\caption{(Color online)
Correlation functions for $C_1=0.3$ and $C^b_3=0.8$ (left),
$C_1=0.3$ and $C^b_3=1.5$ (right).
At $C_1=0.3$ and $C^b_3=0.8$, only the BEC of $a$-atom forms.
On the other hand at $C_1=0.3$ and $C^b_3=1.5$, the FM as well as 
the BECs of $a$ and $b$-atoms form.
This state is denoted as FM+2SF.
}
\label{Corr1}
\end{center}
\end{figure}

Some of the correlation functions that were used for the identification of each phase
are shown in Fig.\ref{Corr1}.
The obtained phase diagram should be compared with that of the case
$t_a=t_b$\cite{BtJ,BtJ1}.
As the result of the mass difference, the phase with single BEC
appears.

In order to study the low-energy excitations in each phase, an effective field theory,
which is derived by means of the ``Hubbard-Stratonovich" transformation,
is very useful\cite{BtJ1}.
For the case of $t_a=t_b$, the effective field theory was derived and 
the number of the NG bosons was identified\cite{BtJ1}.
Similar manipulation is applicable to the present case straightforwardly.
The action of the effective field theory is given as follows for the case $t_a\neq t_b$
and $J_z=0$, though we shall discuss more general case in Sec. IV,
\begin{eqnarray}
A&=& \int d\tau\Big[\sum_{\alpha=a,b,s, \langle i,j \rangle}
(a_\alpha\Phi^\ast_{\alpha i} \Phi_{\alpha j})  
-{1 \over V_0}\sum_{\alpha=a,b, i}(|\dot{\Phi}_{\alpha i}|^2
+V_0^2 |\Phi_{\alpha i}|^2) \nonumber   \\
&&-{1 \over 2V_0}\sum_{i}(|\dot{\Phi}_{s i}|^2
+4V_0^2 |\Phi_{s i}|^2) 
+\sum_i g (\Phi^\ast_{s i}\Phi_{a i}\Phi^\ast_{b i} +\mbox{c.c.}) 
+\sum_{\alpha=a,b,s, i}\lambda_\alpha |\Phi_{\alpha i}|^4\Big],
\label{A0}
\end{eqnarray}
where $\Phi_{\alpha i} \ (\alpha=s,a,b)$ are collective fields for the FM pseudo-spin,
$a$-atom and $b$-atom, respectively, i.e.,
\begin{equation}
e^{i\Omega_{1i}} \Rightarrow \Phi_{si}, \;
e^{i\Omega_{2i}} \Rightarrow \Phi_{ai}, \;
e^{i\Omega_{3i}} \Rightarrow \Phi_{bi}.
\end{equation}
The effective field theory in Eq.(\ref{A0}) is defined in the continuum imaginary-time.
Then the parameters in the action $A$ in Eq.(\ref{A0}) are given as
$a_1=C_1/\Delta\tau$, $a_2=C^a_3/\Delta\tau, a_3=C^b_3/\Delta\tau$ 
and $g\propto V_0$.
It is seen that qualitative structure of the phase diagram shown in Fig.\ref{PD1} 
is easily obtained from the quadratic, cubic and quartic terms of $\Phi_{\alpha i}$ 
in the action $A$ in Eq.(\ref{A0}).

From the effective field theory Eq.(\ref{A0}),  it is easily proved that 
the number of the NG bosons in the FM, $a$-SF, and FM+2SF are
one, one, and two, respectively.
The cubic-coupling terms $g \Phi^\ast_{s i}\Phi_{a i}\Phi^\ast_{b i}+\mbox{c.c.}$
play an essential role for the number of the NG bosons.
One may expect that there appear three NG bosons in the FM+2SF phase
because three U(1) symmetries in the t-J model are spontaneously broken.
However as we showed in the previous paper\cite{BtJ1}, the U(1) spin rotation in the
space $(S^x, S^y)$ is induced by the U(1) phase rotation of the
operators of the $a$ and $b$-atoms,
and therefore the genuine symmetry of the t-J model is $U(1)\times U(1)$.
In the FM+2SF phase, this $U(1)\times U(1)$ symmetry is spontaneously
broken simultaneously and as a result two NG bosons appear.

It is easily seen that $A$ in Eq.(\ref{A0}) has a ``Lorentz invariance", i.e., 
the conjugate variable of the field $\Phi_{\alpha i}$ is 
$\partial_\tau{\Phi}^\ast_{\alpha i}$.
Therefore it is expected that the Higgs modes, which correspond amplitude
modes of $\Phi_{\alpha i}$, appear as elementary excitations\cite{Higgs}.
This point will be discussed rather in detail in Sec.IV.

\begin{figure}[h]
\begin{center}
\includegraphics[width=8cm]{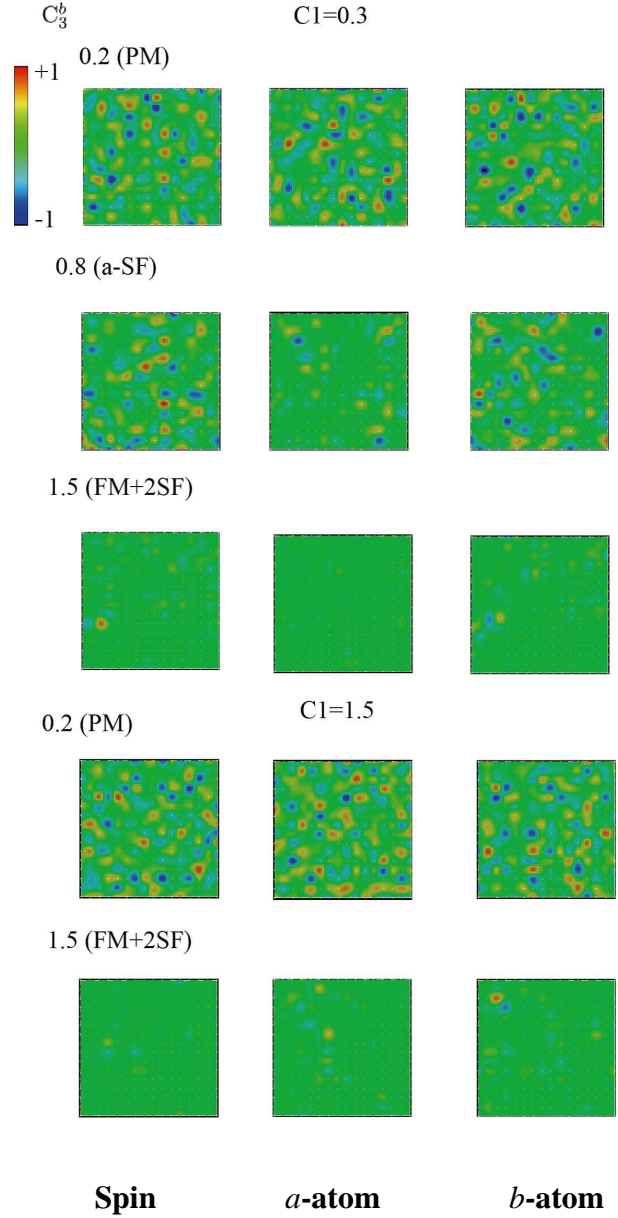} \vspace{0.5cm} \\
\hspace{0.7cm} {\bf Spin}  \hspace{1.5cm} {\bf $a$-atom}
 \hspace{1.5cm} {\bf $b$-atom} 
\caption{(Color online)
Snapshot of vortices for $C_1=0.3$ and $C_1=1.5$.
Spin, $a$-atom and $b$-atom vortices from the left to right columns. 
}
\label{Sp1}
\end{center}
\end{figure}

It is interesting to study topologically stable excitations, i.e., 
vortices, in each phase.
In particular in the FM and SF coexisting phases, one may expect that
vortex for the $xy$-component FM order and that for the SF appear
independently with each other.
However, as the pseudo-spin operator is originally a composite operator
of the $a$-atom and the $b$-atom operators, there might be a close
relation between these vortices. 

Generally vorticity in the $(x-y)$ plane at site $r$, $V_r$, of a complex field
$e^{i\theta_r}$ is defined as,
\begin{eqnarray}
V_r={1 \over 4}\Big[\sin(\theta_{r+\hat{x}}-\theta_r)
+\sin(\theta_{r+\hat{x}+\hat{y}}-\theta_{r+\hat{x}}) 
-\sin(\theta_{r+\hat{x}+\hat{y}}-\theta_{r+\hat{y}})
-\sin(\theta_{r+\hat{y}}-\theta_r)\Big],
\label{Vr}
\end{eqnarray}
where $\hat{x}$($\hat{y}$) is the unit vector in the $x$($y$)-direction. 
By using the definition Eq.(\ref{Vr}), vorticities of the pseudo-spin $S^x_r+iS^y_r$,
$a_r$ and $b_r$ in Eq.(\ref{Sab}) are defined.
We show the numerical calculations of the density of each vorticity in Fig.\ref{Sp1}.
From the phase diagram in Fig.\ref{PD1}, the results in Fig.\ref{Sp1} indicate that
the existence of a long-range order 
obviously suppresses the vortex corresponding to that symmetry.
Careful look at the snapshots reveals that no obvious correlations
between locations of the three type of vortices exist
even though $\Omega_{1r}-\Omega_{2r}+\Omega_{3r}=0$.
It also seems that a solid-like order of vortices does not exist in the disordered
phases, whereas in the ordered phases the density of vortices is very low.
Later we will see that this is in a sharp contrast to the case of the
system in an external magnetic field.

\section{Supersolid}
\setcounter{equation}{0}

Supersolid (SS) is one of the most interesting phenomenon that is expected
to be observed in the cold-atom system.
The SS has both the solid order, which is observed by the density profile, and
the superfluidity.
In this section, we focus on the effect of the $J_z$-term in the Hamiltonian
$H_{\rm tJ}$ in Eq.(\ref{HtJ}) and investigate the possibility of the SS state
as the $J_z$-term enhances Ising like solid order.
To this end, the
internal energy of the system is calculated as a function of
the density deference of $a$ and $b$-atoms in the even-odd sublattices.
Parameter region of the SS state in the phase diagram is clarified by the 
calculation of the internal energy and Bose correlation.

In the practical calculation, we fix $N=1$ and the average density of hole at
each site is put to $30\%$.
We consider the case $J_z>0$ and assume the checkerboard symmetry for the SS
if it exists.
Therefore the
density of the $a$-atom on the even site is equal to that of the $b$-atom
on the odd site and is denoted by $\rho_e$, whereas that of the $a$-atom
on the odd site (the $b$-atom on the even site) $\rho_o=0.7-\rho_e$.
We assume without the loss of the generality $\rho_e\geq \rho_o$,
and define the difference $\Delta\rho=\rho_e-\rho_o\in [0, 0.7]$.
We calculate the internal energy $U$ of the system $H_{\rm tJ}$ in Eq.(\ref{HtJ}) 
by using both the MC simulation and the MF level approximation as,
\begin{eqnarray}
U=\Big\langle 
-\sum_{\langle i,j\rangle} (t_a a^\dagger_{i}a_j
+t_b b^\dagger_{i}b_j+\mbox{h.c.})  
-J\sum_{\langle i,j\rangle}(S^x_{i}S^x_j+S^y_{i}S^y_j)
\Big\rangle
+J_z\sum_{\langle i,j\rangle}S^z_{i}S^z_j,
\label{U}
\end{eqnarray}
where the quantities $\langle \cdots \rangle$ are calculated by the 
MC simulation with the qXY action in Eq.(\ref{AL1}) 
(please notice that the parameters
$C_1$ etc vary as a function of $\Delta\rho$), 
whereas
the last term on the RHS of Eq.(\ref{U}) is evaluated 
by substituting $\rho_e$ and $\rho_o$ and ignoring quantum 
fluctuations.
Then $U$ is obtained as a function $\Delta\rho$ with the other
parameters $t_a$ etc fixed and if $U$ has a 
minimum at {\em nonvanishing} $\Delta\rho$, we conclude that the inhomogeneous 
state with the checkerboard pattern forms.
Existence of the SF is examined by calculating the correlation functions
of the $a$ and $b$-atoms.

\begin{figure}[h]
\begin{center}
\includegraphics[width=5cm]{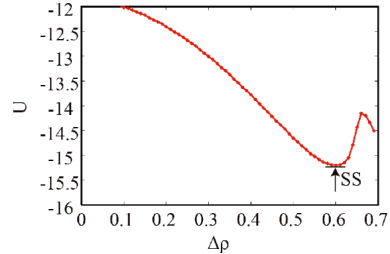}
\vspace{-0.3cm}
\caption{(Color online)
Internal energy $U$ as a function $\Delta\rho$.
U has the absolute minimum at $\Delta\rho\simeq 0.59$.
$J\Delta\tau=(t_a/2)\Delta\tau=(t_b/2)\Delta\tau=30$ and $J_z\Delta\tau=16$.
(See Fig.\ref{SSPD1}.)
}
\label{USS}
\end{center}
\end{figure}

In Fig.\ref{USS}, we show a typical behavior of $U$ as a function $\Delta\rho$.
The state of $\Delta\rho=0$ corresponds to the homogeneous state,
whereas the pure checkerboard configuration of the $a$ and $b$-atoms
corresponds to $\Delta\rho=0.7$ as the average hole density = $30\%$.
From Fig.\ref{USS}, we can see that $U$ generally has three local minima 
for an intermediate value of $J_z$, i.e., $\Delta\rho=0, \Delta\rho_c$ and 
$\Delta\rho=0.7$.
As the value of $J_z$ is increased gradually from zero, the location of 
the absolute minimum of 
$U$ shifts from $\Delta\rho=0$ to $\Delta\rho=\Delta\rho_c(\neq 0, \neq 0.7)$
and finally $\Delta\rho=0.7$.
This behavior comes from the fact that the increase of $\Delta\rho$ 
makes the energy of the hopping term and the $J$-term increase, whereas
the energy of the $J_z$-term decrease.
The SS forms for the parameter region in which 
the absolute minimum of $U$ is located at $\Delta\rho_c(\neq 0, \neq 0.7)$
and the BEC is realized simultaneously.

\begin{figure}[h]
\begin{center}
\includegraphics[width=7cm]{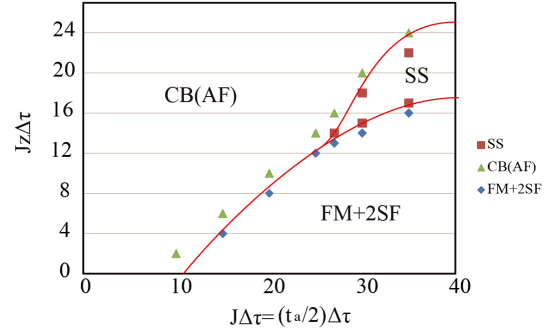}
\vspace{-0.3cm}
\caption{(Color online)
Phase diagram in the $J=t_a$ plain for $J_z>0$ and $t_a=t_b$.
CB(AF) stands for the antiferromagnetic state of the pseudo-spin
(checkerboard state), SS for the supersolid, and FM+2SF for
the SF of the both $a$ and $b$-atoms.
Symbols indicate the location of the phase boundaries verified by the
numerical methods explained in the text.
}
\label{SSPD1}
\end{center}
\end{figure}
\begin{figure}[h]
\begin{center}
\includegraphics[width=8cm]{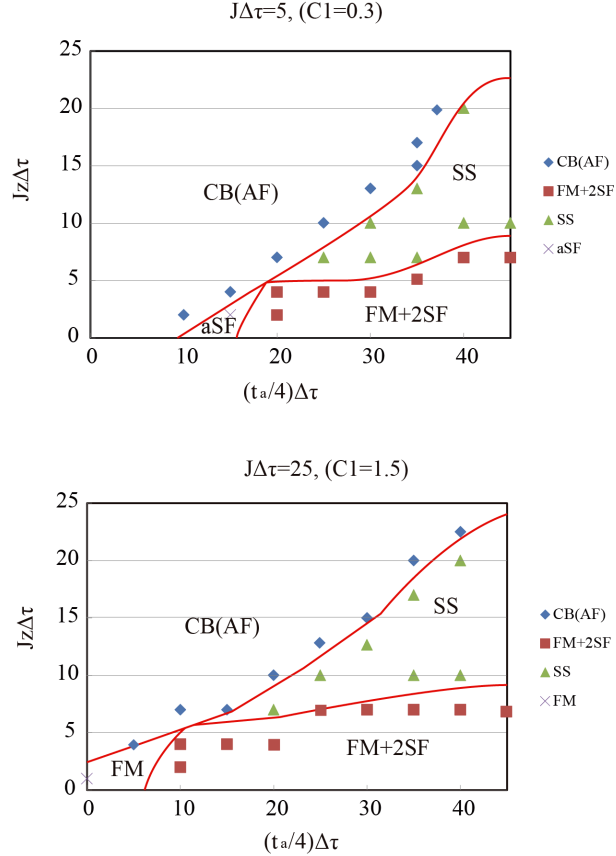}
\vspace{-0.3cm}
\caption{(Color online)
Phase diagram for $J_z>0$ and $t_a=2t_b$.
$J\Delta\tau=25$, $C_1=0.3$ (upper panel) and $C_1=1.5$ (lower panel).
}
\vspace{-1cm}
\label{SSPD2}
\end{center}
\end{figure}

In the practical calculation, we start with FM+2SF states at $J_z=0$
and then increase value of $J_z$.
The internal energy $U$ is calculated as a function of $\Delta\rho$
and then see if the SS forms.
In Figs.\ref{SSPD1} and \ref{SSPD2}, we show the obtained phase diagrams for 
the positive $J_z$.
The SS forms in small parameter regions of the phase diagram.
This result, in particular the phase diagram in Fig.\ref{SSPD2}, 
should be compared with that of the Bose-Hubbard model
of the two-component hard-core boson with a mass difference that was
obtained in Ref.\cite{sansone}. 
It was found there that at half-integer filling factor for each component,
the SS (checkerboard symmetry+SF) forms in the parameter region 
of the strong asymmetric hopping, e.g., $t_a\gg t_b$ and 
$2zt_a/U_{ab} >2$, where $U_{ab}$ is the inter-species repulsion and $z$
is the number of links emanating from a single site.
In the phase diagram obtained in Ref.\cite{sansone}, the SS has the phase 
boundary with the AF phase and 2SF+FM phase as in Figs.\ref{SSPD1} and \ref{SSPD2}.
As the strong asymmetric hopping in the Bose-Hubbard model 
means $J_z\gg J$ in the bosonic t-J model, the phase
diagram obtained in this section is in agreement with that of Ref.\cite{sansone}.

\section{Effective field theory}
\setcounter{equation}{0}
\subsection{Derivation of effective field theory}

Gapless modes in the various phases can be examined by deriving an effective
field theory for low-energy excitations.
Effective field theory was obtained in the previous paper\cite{BtJ1}
for the FM and SF states in the system with $J_z=0$.
By means of the effective field theory, the FM and SF phase transitions
and the low-energy excitations in these phases were studied in detail.
It was verified that the phase diagram of the effective field theory
is in good agreement with that obtained by the MC simulations.

In this section, we shall take into account the effect of the $J_z$-term 
in the Hamiltonian, and derive the effective field theory that can describe the SS.
To this end, we reexamine the amplitude integration of the $a$ and $b$-atoms 
in the path-integral formalism of the system $H_{\rm tJ}+H_V$.
To make the presentation clearer, we fix the gauge such that $\omega_{3r}=0$
as the system $H_{\rm tJ}+H_V$ in the slave-particle representation
is invariant under a local gauge transformation.
The same action is obtained directly by representing the boson fields of the
$a$ and $b$-atoms in terms of their amplitude and phase and using 
the assumption that the average hole density is homogeneous.
Effective field theory in this section is also applicable for the Bose-Hubbard-J
(BHJ) model as the local number constraint becomes irrelevant if $N$ is large enough
and the average hole density is fairly large.

We start with the BHJ Hamiltonian $H_{\rm tJ}+H_V$ in Eqs.(\ref{HtJ})
and (\ref{HV}).
For simplicity, we set $t_a=t_b=t$.
We employ the path-integral representation of the partition function
and use the following parametorization for the $a$ and $b$-bosons,
\begin{eqnarray}
&& a_i=\sqrt{\rho_{ai}+\delta\rho_{ai}}e^{i\phi_{ai}},  \nonumber \\
&& b_i=\sqrt{\rho_{bi}+\delta\rho_{bi}}e^{i\phi_{bi}},
\label{parameter}
\end{eqnarray}
where $\rho_{ai}$ and $\rho_{bi}$ are average number of the $a$ and $b$-atoms
at site $i$, and $\delta\rho_{ai}$ and $\delta\rho_{bi}$ are their fluctuations.
By substituting Eq.(\ref{parameter}) into $H_{\rm tJ}+H_V$,
the hopping term is expressed as follows in the leading order of the average
particle number,
\begin{equation}
-t\sqrt{\rho_{ai}\rho_{aj}} \ e^{-i\phi_{ai}}e^{i\phi_{aj}}+\cdots.
\label{hop}
\end{equation}
Similarly the $xy$-spin term is given as 
\begin{equation}
-J\sqrt{\rho_{ai}\rho_{aj}\rho_{bi}\rho_{bj}}
\ e^{-i\phi_{ai}}e^{i\phi_{aj}} e^{i\phi_{bi}}e^{-i\phi_{bj}}
+\mbox {c.c.}
\label{xyspin}
\end{equation}
In the path integral, the above terms including $e^{i\phi_{\alpha i}}$
are expressed by introducing source terms as
\begin{eqnarray}
\exp \int d\tau\Big[Ce^{-i\phi_{ai}}e^{i\phi_{aj}}\Big]  
=e^{\int d\tau[C{\delta \over \bar{\eta}_{ai}} 
{\delta \over {\eta}_{aj}}]}\cdot
e^{\int d\tau (\eta_{ai}e^{i\phi_{aj}}+\bar{\eta}_{ai}e^{-i\phi_{a i}})},
\label{source}
\end{eqnarray}
for an arbitrary constant $C$.
Similarly for the $xy$-spin composite field $(e^{-i\phi_{a i}}e^{i\phi_{b i}})$,
\begin{eqnarray}
\exp \int d\tau\Big[Ce^{-i\phi_{ai}}e^{i\phi_{aj}}
e^{i\phi_{bi}}e^{-i\phi_{bj}}\Big]   
= e^{\int d\tau[C{\delta \over \bar{\eta}_{si}} 
{\delta \over {\eta}_{sj}}]}\cdot
e^{\int d\tau (\eta_{sj}e^{i\phi_{aj}}e^{-i\phi_{bj}}
+\bar{\eta}_{si}e^{-i\phi_{a i}}e^{i\phi_{bi}})}.
\label{source2}
\end{eqnarray}
On the other hand, the $J_z$-term contains $\delta\rho_{ai}$
and $\delta\rho_{bi}$, $J_z(\delta\rho_{ai},\delta\rho_{bj})$,
and it is expressed as follows by introducing sources $J_{ai}$ and $J_{bi}$,
\begin{eqnarray}
 e^{\int d\tau J_z(\delta\rho_{ai},\delta\rho_{bj})}    
=e^{\int d\tau J_z({\delta \over i\delta J_{ai}},{\delta \over i\delta J_{bj}})} 
\cdot e^{i\int d\tau (\delta\rho_{ai}J_{ai}+\delta\rho_{bi}J_{bi})}.
\label{Jz}
\end{eqnarray}
Finally, the Berry phase and the $V_0$-term are given as 
\begin{eqnarray}
e^{-\int d\tau \sum_{i,\alpha=a,b}(\alpha^\ast_{i}\dot{\alpha}_{i} 
+V_0\delta\rho^2_{\alpha i})}   
=e^{-\int d\tau \sum_{i,\alpha=a,b}(i\delta\rho_{\alpha i}\dot{\phi}_{\alpha i}
+V_0\delta\rho^2_{\alpha i})}.
\label{Berry}
\end{eqnarray}

The final expression in Eqs.(\ref{Jz}) and (\ref{Berry}) is 
a summation of  the linear 
and quadratic terms of $\delta\rho_{\alpha i}$ and then its path integral can be
performed without any difficulty,
\begin{eqnarray}
 \int [D\delta\rho]
e^{i\int d\tau \sum_{i,\alpha}\delta\rho_{\alpha i}J_{\alpha i}}\cdot
e^{-\int d\tau \sum_{i,\alpha}(i\delta\rho_{\alpha i}\dot{\phi}_{\alpha i}
+V_0\delta\rho^2_{\alpha i})}   
=e^{-{1 \over V_0}\int d\tau\sum_{i,\alpha}
(\dot{\phi}_{\alpha i}-J_{\alpha i})^2}.
\label{dotphi}
\end{eqnarray}
By using Eq.(\ref{dotphi}), the path integral of $\phi_{\alpha i}$ can be
performed as follows,
\begin{eqnarray}
\int [D\phi]e^{-{1 \over V_0}\int d\tau
(\dot{\phi}_{\alpha i}-J_{\alpha i})^2}\cdot
e^{\int d\tau (\eta_{\alpha i}e^{i\phi_{\alpha i}}+
\bar{\eta}_{\alpha i}e^{-i\phi_{\alpha i}})}  
=e^{\int d\tau \int d\tau' e^{-V_0|\tau-\tau'|
-i\int^\tau_{\tau'} d\tau''J_{\alpha i}(\tau'')}
\bar{\eta}_{\alpha i}(\tau)\eta_{\alpha i}(\tau')}.
\label{intphi}
\end{eqnarray}
It is not difficult to show that the RHS of  Eq.(\ref{intphi})
can be expressed as a path integral of auxiliary boson fields
$\Phi_{\alpha i}(\tau)$ ($\alpha=a,b$),
\begin{eqnarray}
e^{\int d\tau \int d\tau' e^{-V_0|\tau-\tau'|
-i\int^\tau_{\tau'} d\tau''J_{\alpha i}(\tau'')}
\bar{\eta}_{\alpha i}(\tau)\eta_{\alpha i}(\tau')}  
&=&\int [D\Phi]\exp\Big[ - {1 \over V_0}
\int d\tau \; \Phi^\ast_{\alpha i}(-(\partial_\tau -i {J_{\alpha i}})^2+
V^2_0)\Phi_{\alpha i} \nonumber  \\
&&+\int d\tau(\eta_{\alpha i}\Phi_{\alpha i}+
\bar{\eta}_{\alpha i}\Phi^\ast_{\alpha i})\Big].
\label{Phi}
\end{eqnarray}
Similarly for the $xy$-spin composite field of $\alpha=s$, 
\begin{eqnarray}
\int [D\Phi]\exp\Big[ - {1 \over 2V_0}
\int d\tau \; \Phi^\ast_{s i}(-(\partial_\tau -i {J_{s i}})^2
+4V^2_0)\Phi_{s i}    
+\int d\tau(\eta_{s i}\Phi_{s i}+\bar{\eta}_{s i}\Phi^\ast_{s i})\Big],
\label{Phi2}
\end{eqnarray}
where $J_{si}\equiv J_{ai}-J_{bi}$.

By using the above manipulation, the functional derivatives with respect to 
$\eta_{\alpha i}$ and $J_{\alpha i}$ can be performed straightforwardly
and then the partition function of the BHJ model
is expressed as follows by the path integral of the collective field $\Phi_{\alpha i}$,
\begin{equation}
Z=\int [D\Phi]\ e^{A_\Phi},
\label{ZPhi2}
\end{equation}
\begin{eqnarray}
A_{\Phi} &=&\int d\tau\Biggl[\sum_{\langle i,j\rangle}
\Big[C_{a}\Phi^{\ast}_{ai}\Phi_{aj}
+C_{b}\Phi^{\ast}_{bi}\Phi_{bj}
+C_{s}\Phi^{\ast}_{si}\Phi_{sj}  
-{J_z\over 4}(\rho_{ai}-\rho_{bi})(\rho_{aj}-\rho_{bj})\Big]  \nonumber\\
&+&\sum_{\langle i,j\rangle}\Big[J^a_{1z}\biggl(-\frac{i}{2V_{0}}\Phi^{\ast}_{ai}
\partial^{\leftrightarrow }_{\tau}\Phi_{ai}
-\frac{i}{4V_{0}}\Phi^{\ast}_{si}
\partial^{\leftrightarrow }_{\tau}\Phi_{si}\biggr)  
\times\Big(i\rightarrow  j\Big)  \nonumber\\
&+&J^b_{1z}\biggl(-\frac{i}{2V_{0}}\Phi^{\ast}_{bi}\partial^{\leftrightarrow }_{\tau}\Phi_{bi}-\frac{i}{4V_{0}}\Phi^{\ast}_{si}
\partial^{\leftrightarrow }_{\tau}\Phi_{si}\biggr)  
\times\Big(i\rightarrow j\Big)\nonumber\\
&+&J_{2z}\biggl(-\frac{i}{2V_{0}}\Phi^{\ast}_{ai}
\partial^{\leftrightarrow }_{\tau}\Phi_{ai}-\frac{i}{4V_{0}}\Phi^{\ast}_{si}\partial^{\leftrightarrow }_{\tau}\Phi_{si}\biggr) 
\times\Big(i\rightarrow j,a\rightarrow b\Big)\Big]\nonumber\\
&-&\frac{1}{V_{0}}\sum_{\alpha=a,b,i}\Phi^{\ast}_{\alpha i}(-\partial_{\tau}^{2}+V_{0}^{2})\Phi_{\alpha i}   
-\frac{1}{2V_{0}}\sum_{i}\Phi^{\ast}_{si}
(-\partial_{\tau}^{2}+4V_{0}^{2})\Phi_{si}\nonumber\\
&+&g\sum_{i}(\Phi_{ai}\Phi^\ast_{bi}\Phi^{\ast}_{si}+\mbox{c.c})
-\sum_{\alpha, i}\lambda_\alpha |\Phi_{\alpha i}|^4\Biggr],
\label{actionPhi}
\end{eqnarray}
where
\begin{eqnarray}
 f\partial_\tau^{\leftrightarrow}h&=&
f\partial_\tau h-\partial_\tau f\cdot h, \nonumber \\
 g &\propto& V_{0}, \nonumber \\
 C_{a}&=&t_a\sqrt{\rho_{ai}\rho_{aj}}, \nonumber \\
 C_{b}&=&t_b\sqrt{\rho_{bi}\rho_{bj}},  \\
 C_{s}&=&J\sqrt{\rho_{ai}\rho_{aj}\rho_{bi}\rho_{bj}},  \nonumber \\
J^\alpha_{1z}&=&J_{z}\sqrt{\rho_{\alpha i}\rho_{\alpha j}},
\;\;  \alpha=a,b \nonumber \\
 J_{2z}&=&-2J_{z}\sqrt{\rho_{ai}\rho_{bj}}.\nonumber 
\end{eqnarray}

Existence of the SS can be discussed by using the above effective field theory.
To this end, effective potential of $\Phi_{\alpha}$ and $\Delta\rho$ is
obtained from $A_\Phi$ in Eq.(\ref{actionPhi}) as
\begin{eqnarray}
V(\Phi,\Delta\rho)&=&-\Big[\Big(2dC_{a}-V_0\Big)\Phi^{\ast}_{a}\Phi_{a}
+\Big(2dC_{b}-V_0\Big)\Phi^{\ast}_{b}\Phi_{b} 
+\Big(2dC_{s}-2V_0\Big)\Phi^{\ast}_{s}\Phi_{s}
+{J_z\over 4}(\Delta\rho)^2\Big]  \nonumber \\
&&-g\Big(\Phi_{a}\Phi^\ast_{b}\Phi^{\ast}_{s}+\mbox{c.c}\Big)
+\sum_{\alpha}\lambda_\alpha |\Phi_{\alpha}|^4,
\label{potSS}
\end{eqnarray}
where $d$ is the spatial dimension.
From $V(\Phi,\Delta\rho)$, it is obvious that the $J_z$-term favors
the CB symmetry with $\Delta\rho\neq 0$ 
whereas the hopping terms favor the homogeneous
distribution of the atoms and the BEC with $\Phi_\alpha\neq 0$ 
for sufficiently large $t_a, t_b$ and $J$.
The SS is expected to appear in the region of sufficiently large $J_z$ and 
also the hopping 
amplitude as the result exhibited in Figs.\ref{SSPD1} and \ref{SSPD2}.
On the other hand for the case with $J_z\simeq 0$, it is easily verified
that $V(\Phi,0)$ derives the phase diagram shown in Fig.\ref{PD1}.
In particular by the existence of the cubic term in the potential (\ref{potSS}),
the 2SF state with condensation of $\Phi_a$ and $\Phi_b$ 
accompanies condensation of $\Phi_s$, as seen in the phase diagram in
Fig.\ref{PD1}.


\subsection{Nambu-Goldstone bosons and Higgs modes}

In this subsection, we shall study the low-energy excitations
by using the effective field theory derived in the previous subsection. 
In the following discussion,
we consider the {\em symmetric case} $C_a=C_b, \lambda_a=\lambda_b$ and put
$\langle \Phi_{ai}\rangle=\langle\Phi_{bi} \rangle =v_a$ and 
$\langle \Phi_{si}\rangle =v_s$.
Extension to the SS is straightforward and the similar results are obtained.
Then the effective potential Eq.(\ref{potSS}) with $J_z=0$ reduces to
\begin{eqnarray}
{\cal V}(v_a,v_s)=-4dC_av^2_a+2V_0v^2_a+2\lambda_a v^4_a
-2dC_sv^2_s 
+2V_0v^2_s+\lambda_s v^4_s-2gv^2_av_s.
\label{vvv}
\end{eqnarray}
From Eq.(\ref{vvv}), it is obvious that $v_a\neq 0$ and $v_s \neq 0$
for sufficiently large $C_\alpha (\alpha=a, s)$.
The value of $v_a$ and $v_s$ are obtained by minimizing the potential 
${\cal V}(v_a,v_s)$ with respect to them.
Explicitly, they are solutions to the following a pair of equations,
\begin{eqnarray}
&& 2dC_a+V_0+2v^2_a\lambda_a=gv_s,    \nonumber  \\
&& 2dC_s+2V_0+2v^2_s\lambda_s=g{v^2_a \over v_s}.
\label{eqv}
\end{eqnarray}
The above equations (\ref{eqv}) will be used afterwards to obtain the mass
matrix of the NG bosons and Higgs bosons.

We focus on the case of $v_a\neq 0$ and $v_s \neq 0$, and 
study structure of the massless NG bosons.
To this end, we put $\Phi_\alpha=v_\alpha+i\chi_\alpha$ $(\alpha=a, b, s)$
assuming the positive $v_a=v_b$ and $v_s$ without loss of generality.
We take a continuum description for simplicity.
The mass matrix of $\chi_\alpha$ is obtained as follows from Eq.(\ref{actionPhi}),
\begin{equation}
{\cal M}_{\rm NG}(\chi)=(\chi_a \chi_b \chi_s) \hat{M}_{\rm NG} 
\begin{pmatrix}
\chi_a \\
\chi_b \\
\chi_s
\end{pmatrix},
\label{Vchi}
\end{equation}
where
\begin{equation}
\hat{M}_{\rm NG}=g
\begin{pmatrix}
v_s & v_s & -v_a \\
v_s & v_s & -v_a \\
-v_a & -v_a & {v^2_a \over v_s}
\end{pmatrix}.
\label{Mchi}
\end{equation}
The above mass matrix $\hat{M}_{\rm NG}$ is easily diagonalized by using a unitary
matrix $\hat{U}$ as 
\begin{equation}
\hat{U}^{-1}\hat{M}_{\rm NG}\hat{U}=g
\begin{pmatrix}
0 & 0 & 0 \\
0 & 0 & 0 \\
0 & 0 & {2v^2_s+v^2_a \over v_s}
\end{pmatrix}.
\label{UMU}
\end{equation}
From Eq.(\ref{UMU}), it is obvious that there exist two gapless modes (NG bosons)
and one gapful mode in $\chi_\alpha$.
Explicitly,
\begin{equation}
\begin{pmatrix}
\psi_1 \\
\psi_2 \\
\psi_3
\end{pmatrix}
=
\hat{U}
\begin{pmatrix}
\chi_a \\
\chi_b \\
\chi_s
\end{pmatrix},
\label{psi}
\end{equation}
where $\psi_1$ and $\psi_2$ are NG bosons.

Let us derive the dispersion relation of the above NG bosons.
By substituting  $\Phi_\alpha=v_\alpha+i\chi_\alpha$ $(\alpha=a, b, s)$ into
the action of the effective field theory Eq.(\ref{actionPhi}) and
taking the continuum description,
the time-derivative term of $\chi_\alpha$ has the following structure,
\begin{equation}
{\cal T}=\sum_{\alpha,\beta=a,b,s}
\dot{\chi}_\alpha \hat{T}_{\alpha\beta}\dot{\chi}_\beta,
\label{Tchi}
\end{equation}
where $\hat{T}$ is a matrix.
The gapful mode $\psi_3$ in Eq.(\ref{UMU}) can be safely integrated out in the 
path-integral, and the resultant action of the two NG modes $\psi_1$ and $\psi_2$
has the following form,
\begin{eqnarray}
\sum_{\alpha,\beta=1,2}
(\hat{P}_{\alpha\beta}\partial_\tau{\psi}_\alpha \partial_\tau{\psi}_\beta
+\hat{Q}_{\alpha\beta}\partial^2_\tau{\psi}_\alpha \partial^2_\tau{\psi}_\beta)
+\sum_{\alpha=1,2, \mu=x,y}(\partial_\mu\psi_\alpha)^2,
\label{actionpsi}
\end{eqnarray}
where $\hat{P}$ and $\hat{Q}$ are
matrices.
From Eq.(\ref{actionpsi}), the dispersion relation $\omega({\bf k})$
has the form
\begin{equation}
\omega^2({\bf k}) \propto -f+\sqrt{f^2+{\bf k}^2},
\end{equation}
where $f$ is a real number and $\omega({\bf k})$ has a relativistic dispersion
relation for small ${\bf k}$,
\begin{equation}
\omega({\bf k}) \propto |{\bf k}|.
\label{disp1}
\end{equation}

Let us turn to the Higgs bosons, i.e., the amplitude mode of $\Phi_\alpha$.
As the action Eq.(\ref{actionPhi}) shows, the conjugate field theory of $\Phi_\alpha$
is essentially $\partial_\tau{\Phi}^\ast_\alpha$, and therefore the amplitude and 
phase modes are independent dynamical variables.
In the original bosonic t-J model and also the Bose-Hubbard-J model,
the boson operators, e.g., $\hat{a}_i$ and $\hat{a}^\dagger_i$ are 
conjugate with each other, and this lead to the fact that the amplitude and the phase 
are also conjugate with each other and they are {\em not} independent variables.
The derivation of the effective field theory in this section
eloquently tell us that at low 
energies and close to the phase boundary, the order parameters $\Phi_\alpha$
behave as relativistic fields.
This fact was revealed in the seminal paper Ref.\cite{fisher}.

To study the Higgs modes, we put
\begin{eqnarray}
\Phi_{\alpha i}&=&v_{\alpha}+\eta_{\alpha i} \nonumber\\
\Phi^{\ast}_{\alpha i}&=&v_{\alpha}+\eta_{\alpha i}. 
\label{Higgsfield}
\end{eqnarray}
By substituting Eq.(\ref{Higgsfield}) into Eq.(\ref{actionPhi}),
the mass matrix of the Higgs field $\eta_{\alpha i}$ is obtained as,
\begin{equation}
{\cal M}_{\rm H}=
(\eta_{a},\eta_{b},\eta_{s})\hat{M}_{H}\left( \begin{array}{c}
 \eta_{a}\\
 \eta_{b}\\
 \eta_{s}
\end{array} \right),
\end{equation} 
where
\begin{eqnarray}
\hat{M}_{H}&=&
\begin{pmatrix}
     4v^{2}_{a}\lambda_{a}+gv_{s} & -gv_{s} & -gv_{a} \\
     -gv_{s} & 4v^{2}_{a}\lambda_{a}+gv_{s}  & -gv_{a} \\
      -gv_{a} &  -gv_{a} & 4v^{2}_{s}\lambda_{s}+g\frac{v^{2}_{a}}{v_{s}}
\end{pmatrix}. \nonumber
\end{eqnarray}
The above matrix $\hat{M}_{\rm H}$ has three eigenvalues 
($\lambda_{1}$,$\lambda_{\pm}$),
\begin{eqnarray}
\lambda_{1}=4v^{2}_{a}\lambda_{a}+2gv_s, \;\;
\lambda_{\pm}=\frac{F\pm\sqrt{G}}{2},
\label{Higgsgap}
\end{eqnarray}
\begin{eqnarray}
F&=&4(v^{2}_{a}\lambda_{a}+v^{2}_{s}\lambda_{s})
+g\frac{v^{2}_{a}}{v_{s}}, \nonumber\\
G&=&\biggl(4v^{2}_{a}\lambda_{a}-4v^{2}_{s}\lambda_{s}-g\frac{v^{2}_{a}}{v_{s}}\biggr)^{2}+8g^{2}v^{2}_{a}.
\label{FG}
\end{eqnarray}

It is quite instructive to consider the limit $v_a \rightarrow 0$,
i.e., {\em approaching to the phase boundary of the FM and FM+2SF phases}.
In this limit,
\begin{eqnarray}
\hat{M}_{H}&\rightarrow&
\begin{pmatrix}
     gv_{s} & -gv_{s} & 0 \\
     -gv_{s} & gv_{s}  & 0 \\
      0 &  0 & 4v^{2}_{s}\lambda_{s}
\end{pmatrix}, \nonumber
\end{eqnarray}
and the massgaps become as
\begin{equation}
\lambda_1 \rightarrow 2gv_s, \;\;
\lambda_+ \rightarrow 8v^2_s\lambda_s, \;\;
\lambda_- \rightarrow O(v^2_a) \rightarrow 0.
\label{Higgsgap2}
\end{equation}
In Eq.(\ref{Higgsgap2}), $\lambda_+$ is the ordinary massgap
of the Higgs boson corresponding to the amplitude mode of the spin
degrees of freedom.
On the other hand,
$\lambda_-$, which tends to vanish at the phase boundary, corresponds
to the Higgs boson of the SF amplitude.
As we are considering the phase boundary at which the SFs of both the $a$
and $b$-atoms tend to disappear, one may expect the appearance of 
two Higgs modes with a vanishing massgap, whereas in the present case only one exists.
From Eq.(\ref{Higgsgap2}), it is obvious that the finiteness of $\lambda_1$ 
at the phase boundary results from the cubic coupling in action (\ref{actionPhi}).
This cubic term comes from the fact that the spin operator is a composite operator
of the $a$ and $b$-atoms, and the spin U(1) symmetry is
nothing but the symmetry of the phase rotation of these operators.
Then the number of the NG bosons is two but not three in the FM+2SF phase.
The above behavior of the Higgs bosons in Eq.(\ref{Higgsgap2}) is consistent 
with the number of the NG bosons.

In the experiment of a single component gas like $^{87}$Rb atoms,
the Higgs mode is ambiguously identified by observing softening of spectral
response on approaching to the phase boundary of the superfluid and the Mott 
insulator\cite{Higgs}.
For the two-component Bose gas, above result and Eq.(\ref{Higgsgap2}) 
shows that {\em the behavior of the Higgs model and the number of the NG bosons 
crucially depends on the magnitude of the $J$-coupling}, i.e.,
for sufficiently large $J$ (FM+2SF$\rightarrow$FM transition)
only one softening Higgs mode appears whereas for 
small $J$ (FM+2SF$\rightarrow$PM transition), 
$v_s=0$ and therefore there appear two softening Higgs modes for each
BEC of $a$ and $b$-atoms.
We hope that this phenomenon will be observed by experiment near future.

\section{Ground state in external magnetic fields and vortex lattice}
\setcounter{equation}{0}

In this section, we shall  study the two-component boson system
in an effective external magnetic field.
The artificial magnetic field can be generated in experiment by, e.g., rotating the system 
with a confining potential or laser-assisted tunneling 
method\cite{rotation,magnetic,tung,will,zoller}.
In particular, we are interested in how the Bose-condensed states observed
in the previous section will evolve as the strength of the external magnetic
field is increased.
After investigating this problem, we shall study a BEC system that is closely
related to a single-atom system in a 
staggered external magnetic field, which was recently realized by
experiment\cite{staggered}.

\vspace{-0.5cm}

\subsection{qXY model in a uniform magnetic field} 

System action on the cubic space-time lattice including the effect of 
the magnetic field is given as follows,
\begin{eqnarray}
A_{\rm Lxy}(A)&=&A_{{\rm L}\tau}+
A_{\rm L}(e^{i\Omega_\sigma},e^{-i\Omega_\sigma};A),  \nonumber  \\
A_{\rm L}(e^{i\Omega_\sigma},e^{-i\Omega_\sigma};A) &=& 
-\sum_{\langle r,r'\rangle}
\Big(C^a_3\cos (\Omega_{2r}-\Omega_{2r'}+A^a_{r,r'}) 
+C^b_3\cos (\Omega_{3r}-\Omega_{3r'}+A^b_{r,r'}) \nonumber \\
&&+C_1\sum_{\langle r,r'\rangle}\cos (\Omega_{1r}-\Omega_{1r'}
-A^a_{r,r'}+A^b_{r,r'})\Big),
\label{AL2V}
\end{eqnarray}
where $A^a_{r,r'}$ and $A^b_{r,r'}$ are vector potentials that 
the $a$ and $b$-atoms feel, respectively, and given by
\begin{eqnarray}
 \begin{pmatrix}
A^a_{r,r+\hat{x}}  \\
A^a_{r,r+\hat{y}} 
\end{pmatrix}
=\begin{pmatrix}
\pi f \times y  \\
-\pi f\times x  
\end{pmatrix},  
\hspace{0.5cm}
 \begin{pmatrix}
A^b_{r,r+\hat{x}}  \\
A^b_{r,r+\hat{y}} 
\end{pmatrix}
=\begin{pmatrix}
\pi f' \times y   \\
-\pi f'\times x  
\end{pmatrix},  
\hspace{0.5cm}
 \mbox{otherwise zero},
\label{Alink}
\end{eqnarray}
with $r=(x,y,\tau)$. 
In Eq.(\ref{Alink}), $f$ and $f'$ are the parameters for the strength of 
the magnetic field, i.e., $2\pi f$ ($2\pi f'$) is the magnetic flux per
plaquette for the $a$-atom ($b$-atom).
In the practical calculation, we employed the periodic boundary condition.
Therefore, the values of $f$ and $f'$ are restricted as $\pi f L=2n\pi$ and
$\pi f' L=2n'\pi$ where$L$ is the linear size of the system, and
$n$ and $n'$ are integers.

\begin{figure}[h]
\begin{center}
\includegraphics[width=5cm]{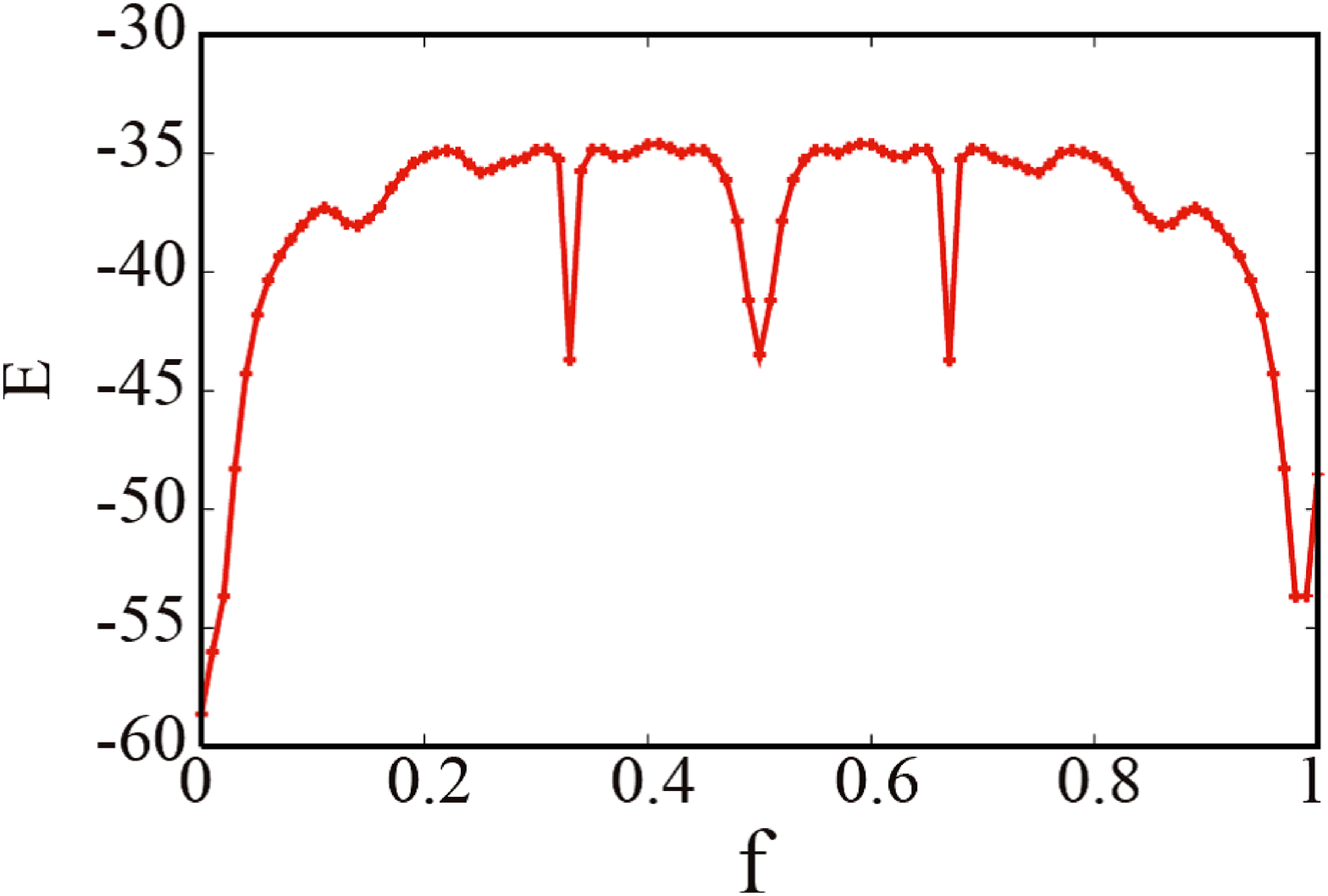}
\includegraphics[width=5cm]{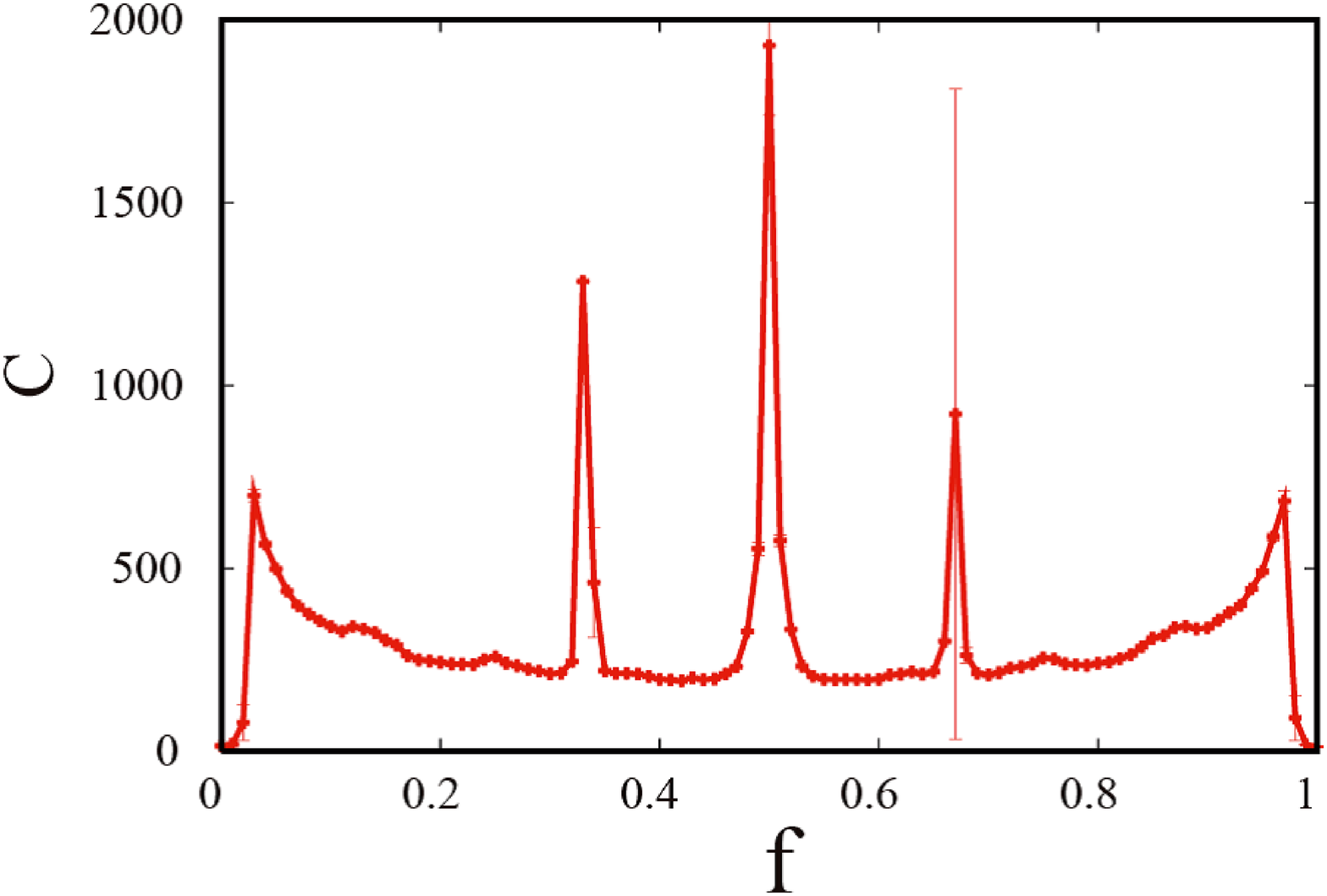}
\vspace{-0.3cm}
\caption{(Color online)
Internal energy $E$ and specific heat $C$ as a function of the magnetic field $f$.
$L=24$.
}
\label{ECmag}
\end{center}
\end{figure}

We first consider the system of the same mass, i.e., $C^a_3=C^b_3$, 
and the state of the FM+2SF in the phase diagram for $J_z=0$, which exists for
sufficiently large $C_1$ and $C^a_3=C^b_3$ as we showed in the previous
paper\cite{BtJ1}.

\begin{figure}[h]
\begin{center}
\includegraphics[width=4cm]{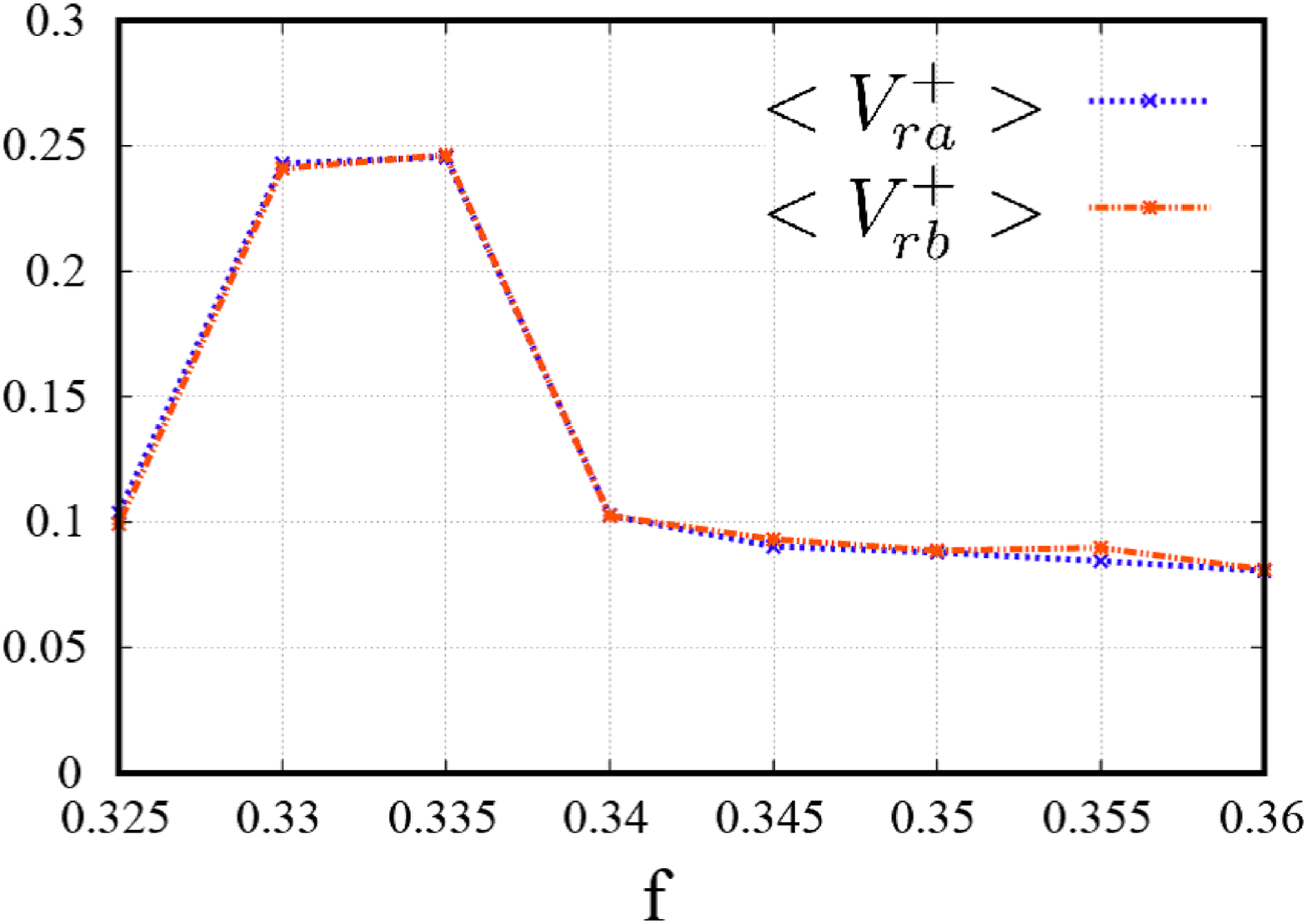}
\includegraphics[width=4cm]{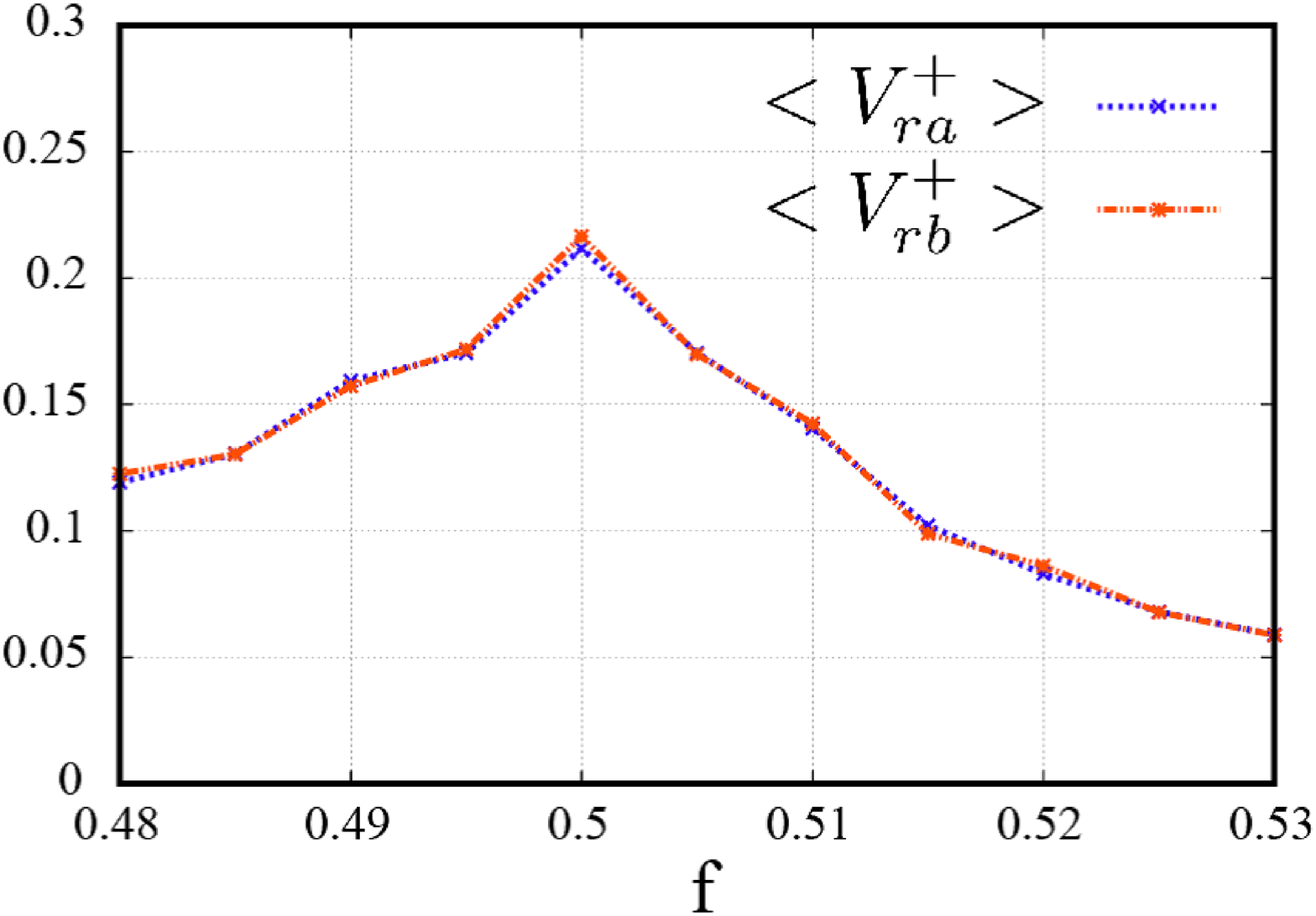}
\includegraphics[width=4cm]{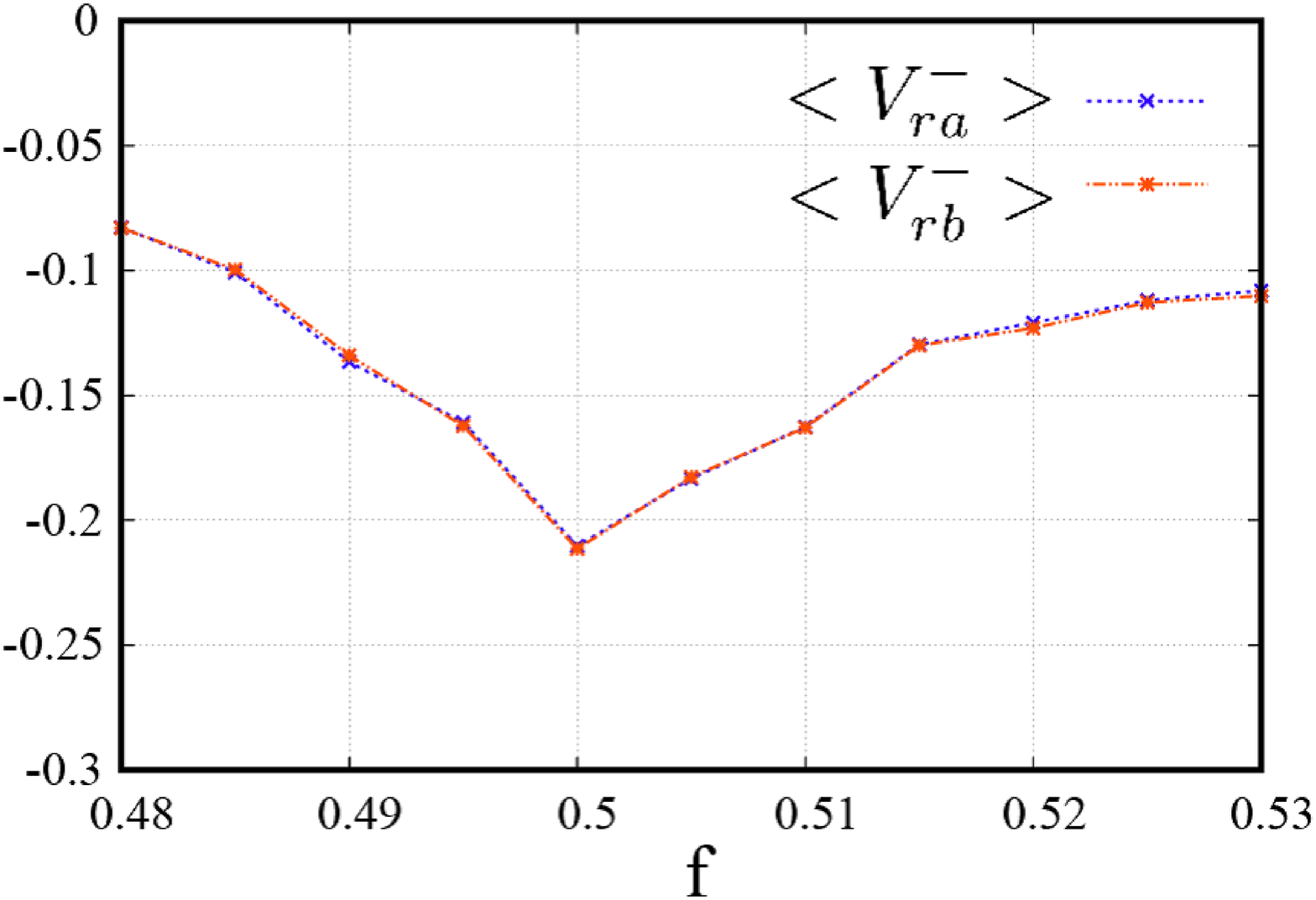}
\includegraphics[width=4cm]{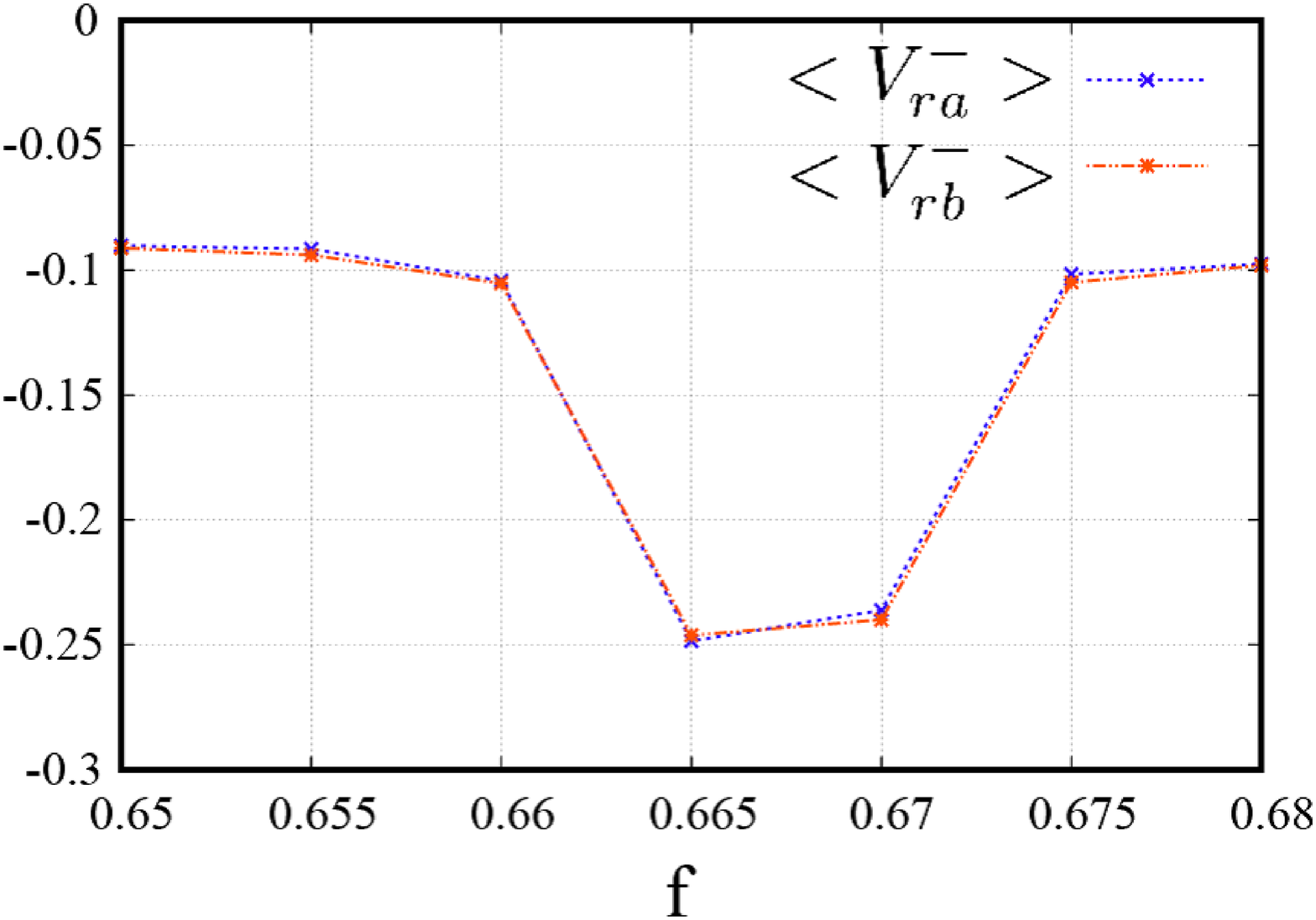}
\vspace{-0.3cm}
\caption{(Color online)
Vortex density as a function the magnetic field $f$.
The results indicate that some specific states form at 
$f=1/3, 1/2$ and $2/3$.
}
\label{vor_density}
\end{center}
\end{figure}

\begin{figure}[h]
\begin{center}
\includegraphics[width=6cm]{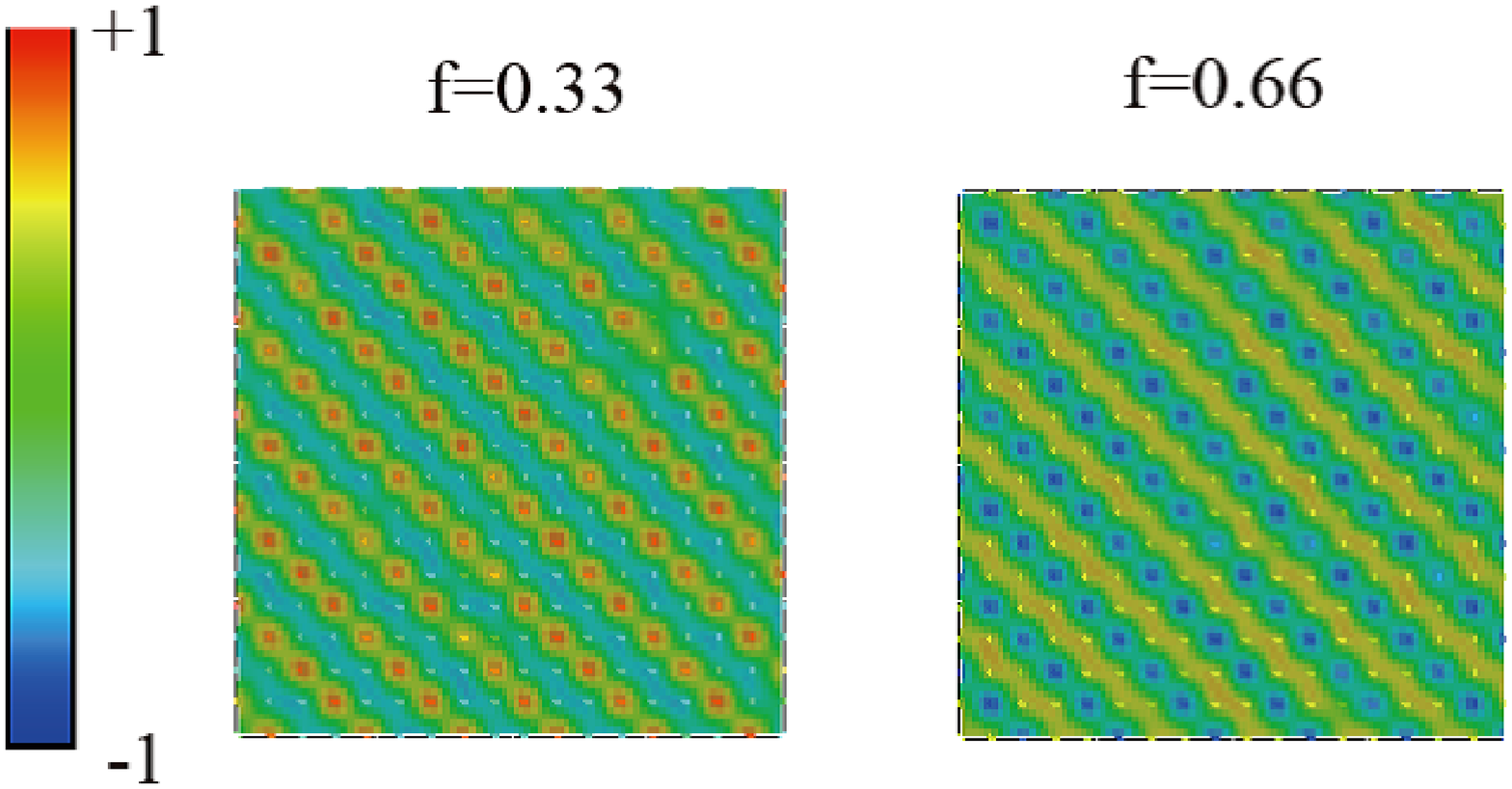}
\includegraphics[width=6cm]{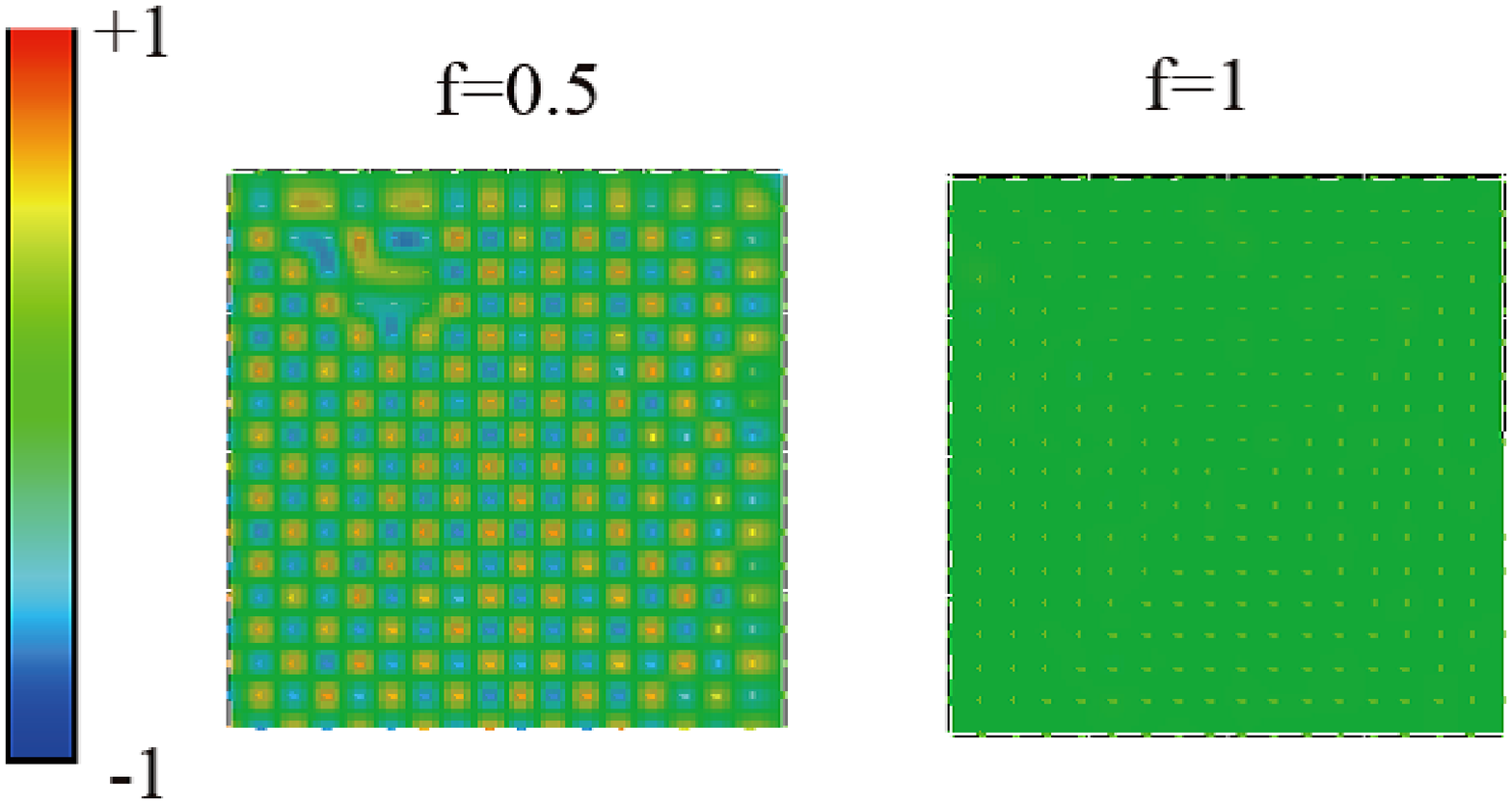}
\vspace{-0.3cm}
\caption{(Color online)
Snapshots of vortex for $f=0.33, 0.66, 0.5$ and $f=1$.
The results indicate the existence of some kind of vortex-solid order
except $f=1$.
See Fig.\ref{con_vortex}.
}
\label{snap_vortex}
\end{center}
\end{figure}

In Fig.\ref{ECmag}, we show the internal energy $E$ and specific heat $C$ 
as a function of $f(=f')$ for $C_1=3$
and $C^a_3=40$ obtained by the MC simulation.
The specific heat shown in Fig.\ref{ECmag} seems to indicate that the BEC is destroyed
quite easily by the external magnetic field with very small value of $f\sim 0.02$.
We verified that the BEC is actually destroyed by calculating the 
boson correlation function.
However, $E$ and $C$ in Fig.\ref{ECmag} 
obviously indicate the existence of  certain stable states for specific values of $f$,
i.e., $f={1 \over 3}, {1 \over 2}$ and ${2\over 3}$\cite{nakano,FXY}.
We also measured the average vortex density $\langle V^+_r\rangle$
and anti-vortex density $\langle V^-_r\rangle$
as a function of $f$, and the result is shown in Fig.\ref{vor_density}
for $f\sim 1/3, 1/2$ and $2/3$,
where the vortex density $V^+_r$ is defined as 
\begin{equation}
V^+_r \equiv\begin{cases}
          V^A_r, & \text{$V^A_r >0.6$}  \\
          0,    & \text{$V^A_r <0.6$},
         \end{cases}
\label{defV}
\end{equation}
with
\begin{eqnarray}
V^A_r&=&{1 \over 4}\Big[\sin(\theta_{r+\hat{x}}-\theta_r-A_{r,r+\hat{x}}) 
+\sin(\theta_{r+\hat{x}+\hat{y}}-\theta_{r+\hat{x}}  
-A_{r+\hat{x},r+\hat{x}+\hat{y}}) \nonumber \\
&&-\sin(\theta_{r+\hat{x}+\hat{y}}-\theta_{r+\hat{y}}
-A_{r+\hat{y},r+\hat{x}+\hat{y}}) 
-\sin(\theta_{r+\hat{y}}-\theta_r-A_{r,r+\hat{y}})\Big],
\label{VAr}
\end{eqnarray}
and similarly for $V^-_r$\cite{cutoff}.
See also snapshots of vortices in Fig.\ref{snap_vortex}, which indicate
the existence of  some kind of vortex lattice for $f=1/3, 1/2$ and $2/3$.


\begin{figure}[h]
\begin{center}
\includegraphics[width=4cm]{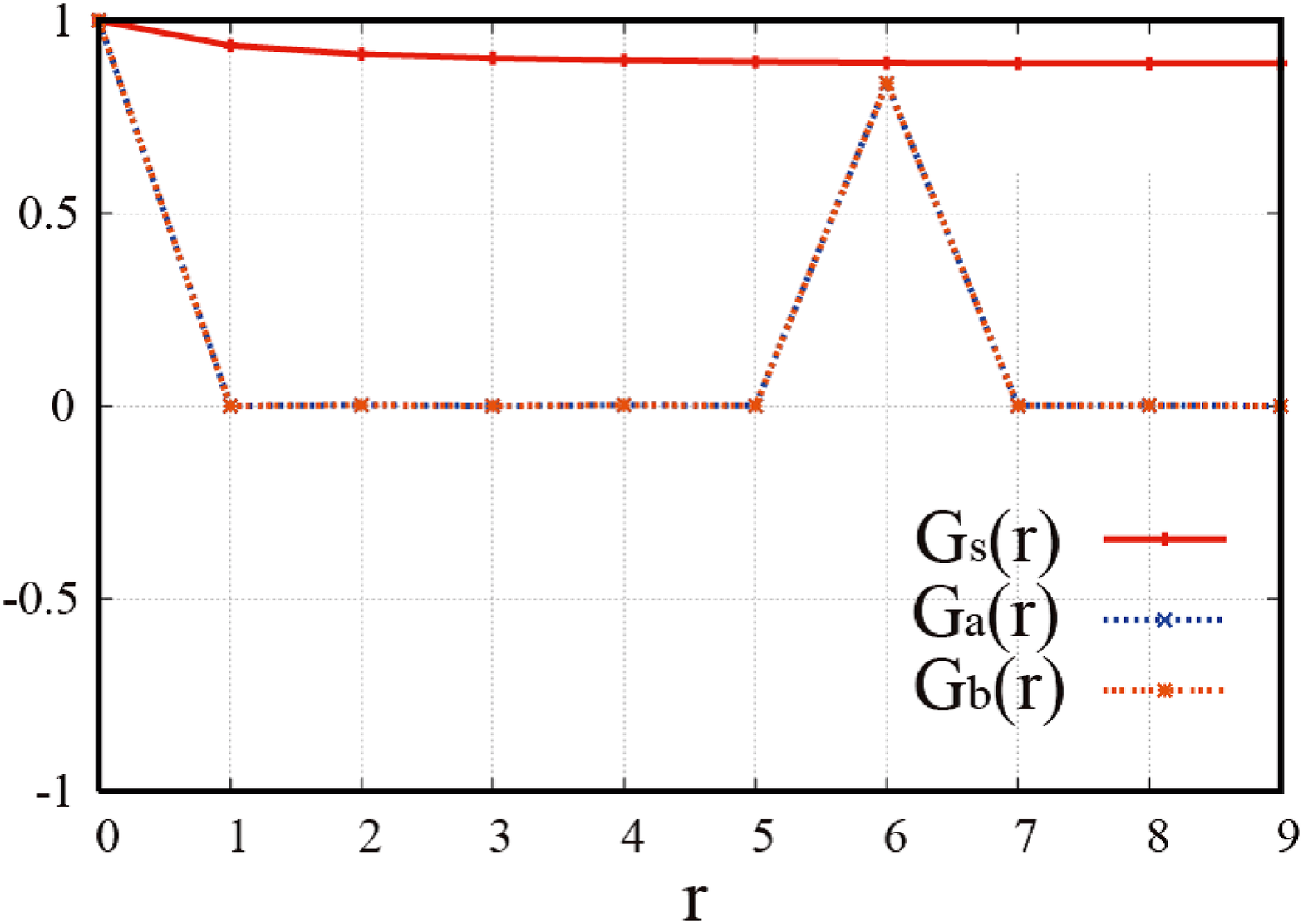}
\includegraphics[width=4cm]{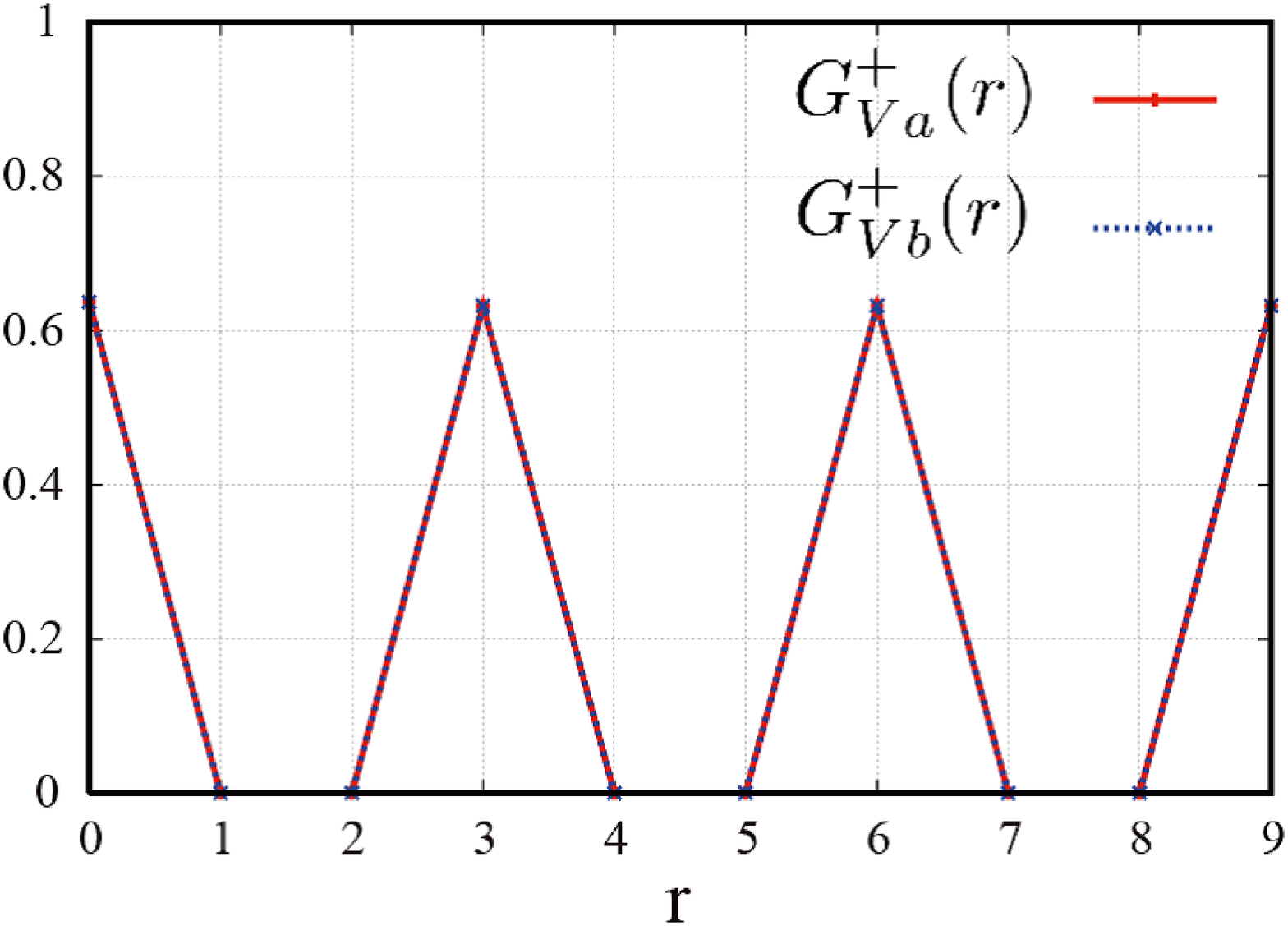}
\vspace{-0.3cm}
\caption{(Color online)
Particle and vortex correlation functions for $f=0.33$ and
$t^a=t^b$.
}
\label{crr_f=033}
\end{center}
\end{figure}
\begin{figure}[h]
\begin{center}
\includegraphics[width=4cm]{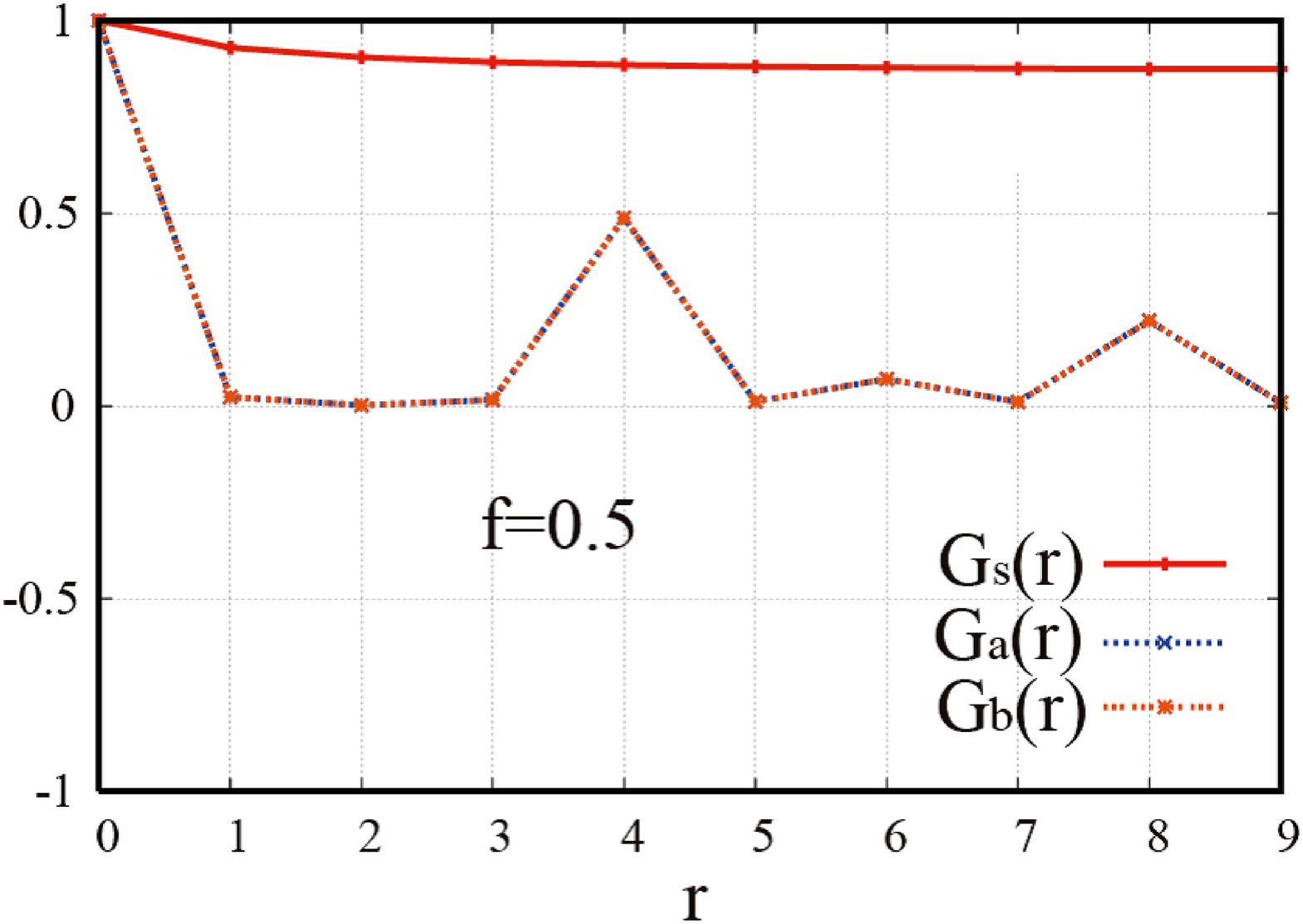}
\includegraphics[width=4cm]{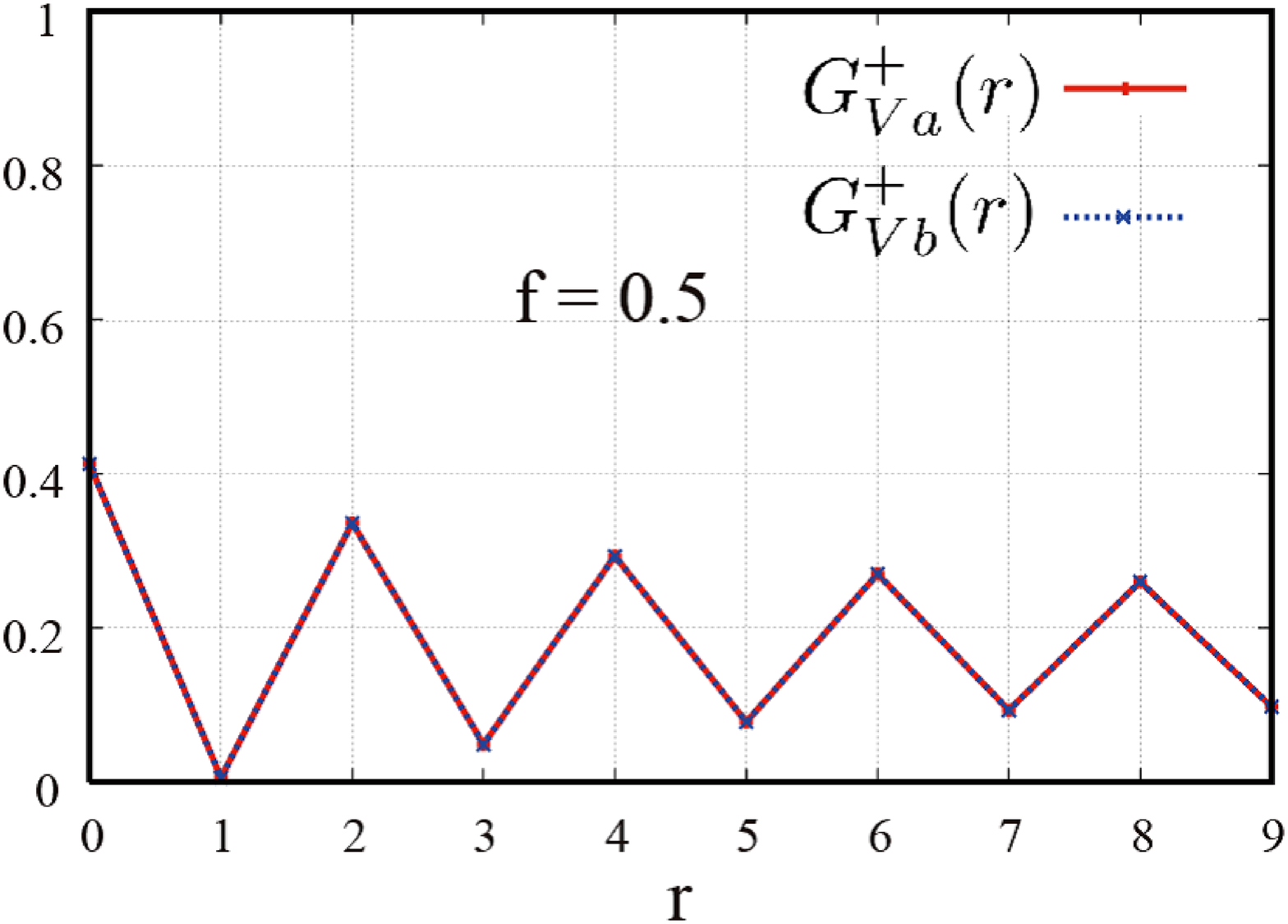}
\includegraphics[width=4cm]{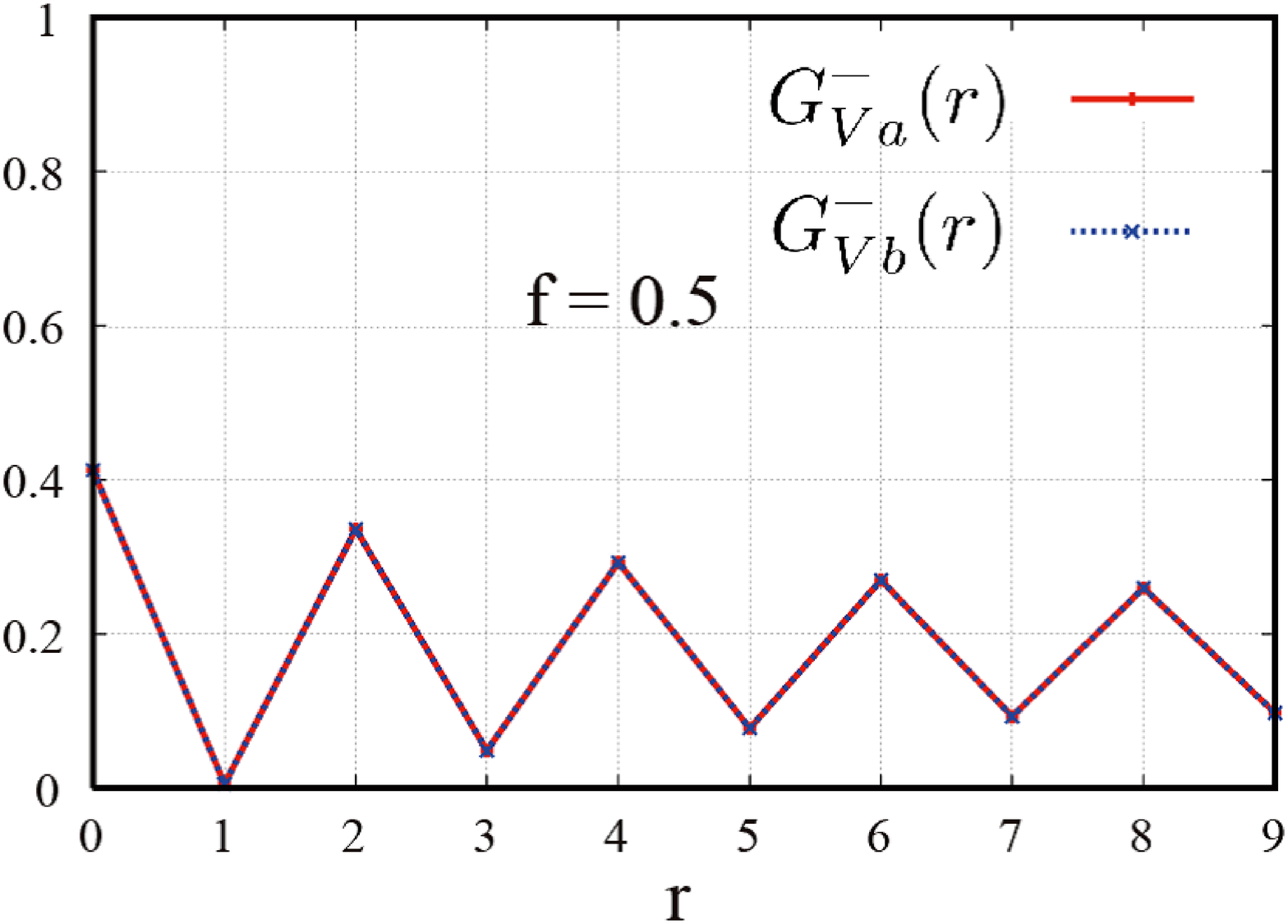}
\includegraphics[width=4cm]{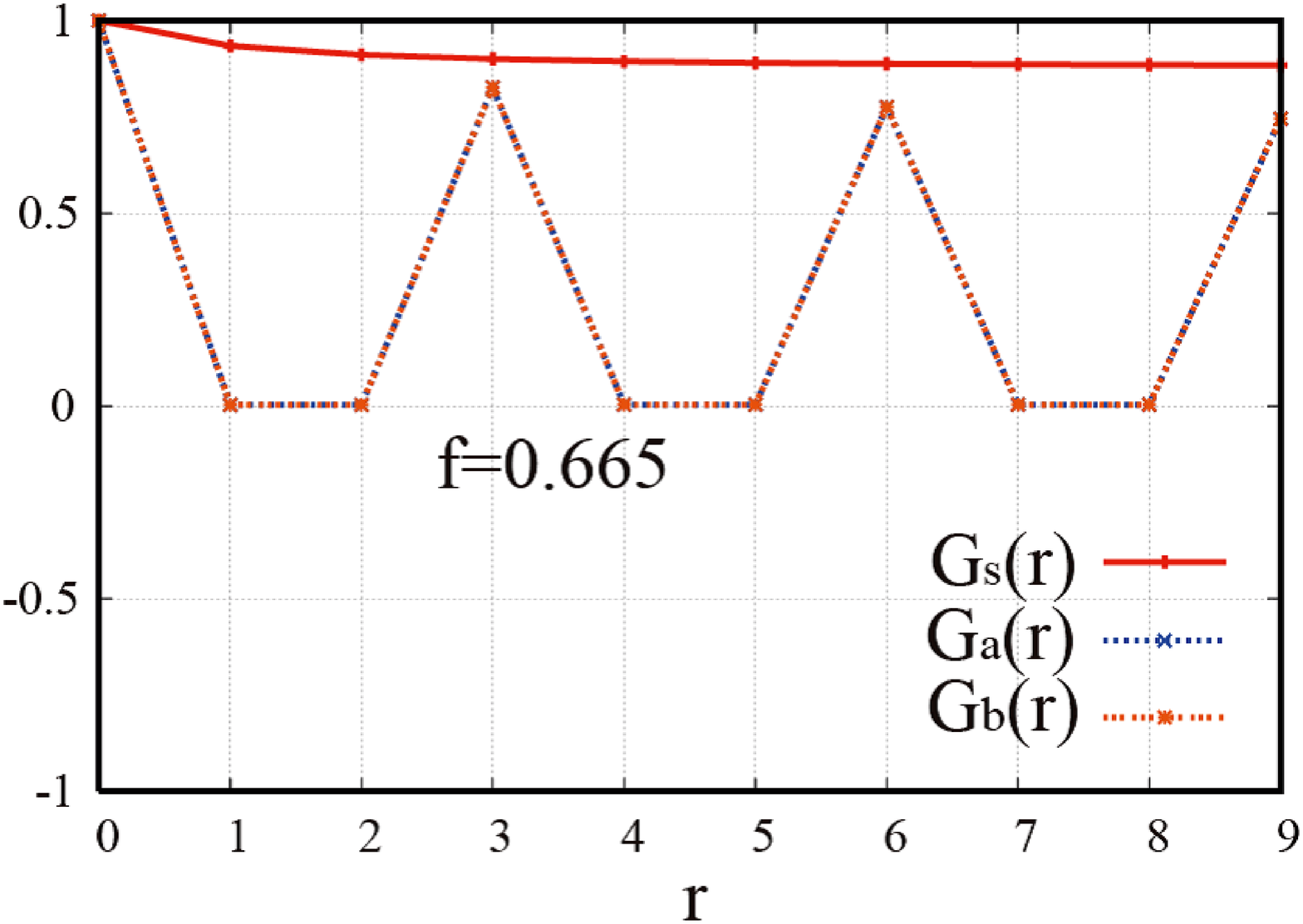}
\includegraphics[width=4cm]{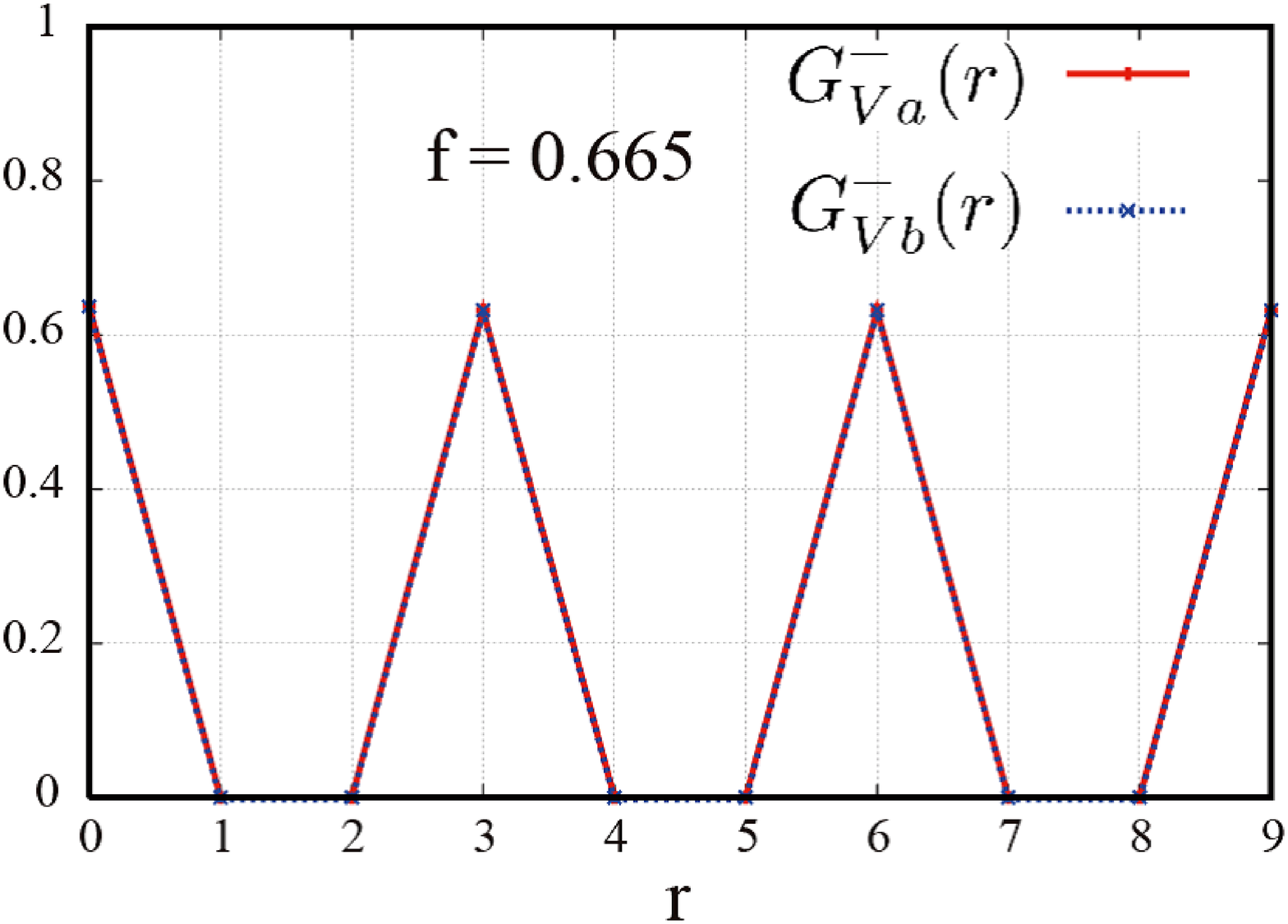}
\vspace{-0.3cm}
\caption{(Color online)
Various correlation functions for $f=0.5$ and $f=0.665$.
$t^a=t^b$.
}
\label{crr_f=067}
\end{center}
\end{figure}

It is useful to observe the correlation functions of the $a,\ b$-atom
and spin for investigating the states appearing at $f=1/3, 1/2$ and $2/3$ in more detail.
These correlation functions for $f=1/3$ are shown in Fig.\ref{crr_f=033}.
The spin correlation has an ordinary LRO, whereas the boson correlation
exhibits a specific spatial pattern.
Similarly the vortex correlations are obtained as in Fig.\ref{crr_f=033},
which also exhibit certain spatial pattern.
Here we define the correlation function of the vortex as
\begin{eqnarray}
G^+_{\rm V}(r) = {1 \over 4}\Big(\langle V^+_{r_0}V^+_{r_0+\hat{x}r} \rangle
+\langle V^+_{r_0}V^+_{r_0-\hat{x}r} \rangle  
+\langle V^+_{r_0}V^+_{r_0+\hat{y}r} \rangle
+\langle V^+_{r_0}V^+_{r_0-\hat{y}r} \rangle\Big),
\end{eqnarray}
where $r_0$ is the location of vortex, i.e., $V^A_{r_0}\simeq 1$.
The correlation function of anti-vortex $G^-_{\rm V}(r)$ is defined similarly,
though it is vanishingly small for $f=1/3$.
It is obvious that the boson and vortex have qualitatively the same
correlations though the boson correlations have a strong correlation
for $r=6$ whereas the vortex ones for $r=3$.
All the above results indicate the existence of certain specific 
configuration of vortices.
To verify this expectation, we also studied the cases $f=1/2$ and $2/3$ in which
a stable state is expected to exist from the result of $E$ in Fig.\ref{ECmag}.
From the results in  Figs.\ref{crr_f=033} and \ref{crr_f=067}, it is obvious that
vortex solid (vortex lattice) forms at these values of $f$ as the snapshots in 
Fig.\ref{snap_vortex} show\cite{vortex_GP}.
We examined other cases from the above values of $f$, and found that
no LROs exist in any correlation.

\begin{figure}[h]
\begin{center}
\includegraphics[width=7cm]{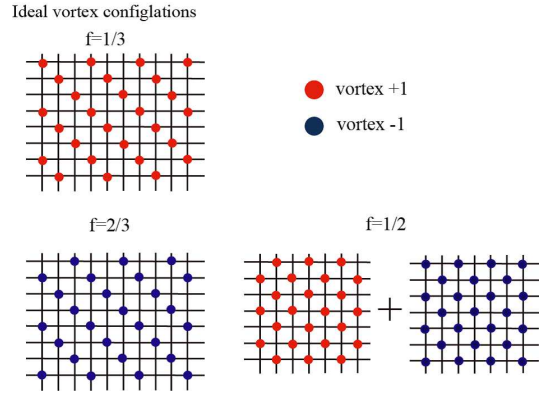}
\vspace{-0.3cm}
\caption{(Color online)
Dominant configuration of vortices for $f=1/3, 1/2$ and $2/3$.
Correlation functions indicate that a state of quantum superposition of vortex 
and anti-vortex is realized at $f=1/2$.
Vortices are located on sites of the {\em dual lattice} of the original square lattice.
}
\label{con_vortex}
\end{center}
\end{figure}

In Fig.\ref{con_vortex}, we show the typical configurations of vortices that are
obtained through a careful look at the vortex snapshots
and the vortex correlation functions.
We also observed by the MC simulation that location of each vortex sightly fluctuates 
from the above through the MC update, which can be understood as a 
quantum fluctuation.
It should be remarked that the lattice spacing of the ``boson lattice"
doubles that of the vortex lattice for $f=1/3$ and $f=1/2$, whereas
for $f=2/3$ they are the same.
This vortex lattice is expected to be observed by the density profile of the BEC.
Furthermore its direct measurement might be possible by using recent experimental
techniques.
This will be discussed in the following subsection.


\begin{figure}[h]
\begin{center}
\includegraphics[width=6cm]{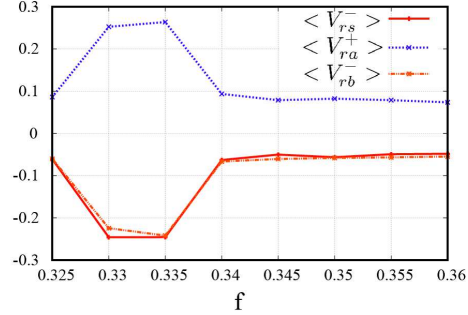}
\vspace{-0.3cm}
\caption{(Color online)
Density of vortices as a function of $f$.
$t_a=t_b/2$.
}
\label{vortex_density_md}
\end{center}
\end{figure}
\begin{figure}[h]
\begin{center}
\includegraphics[width=5cm]{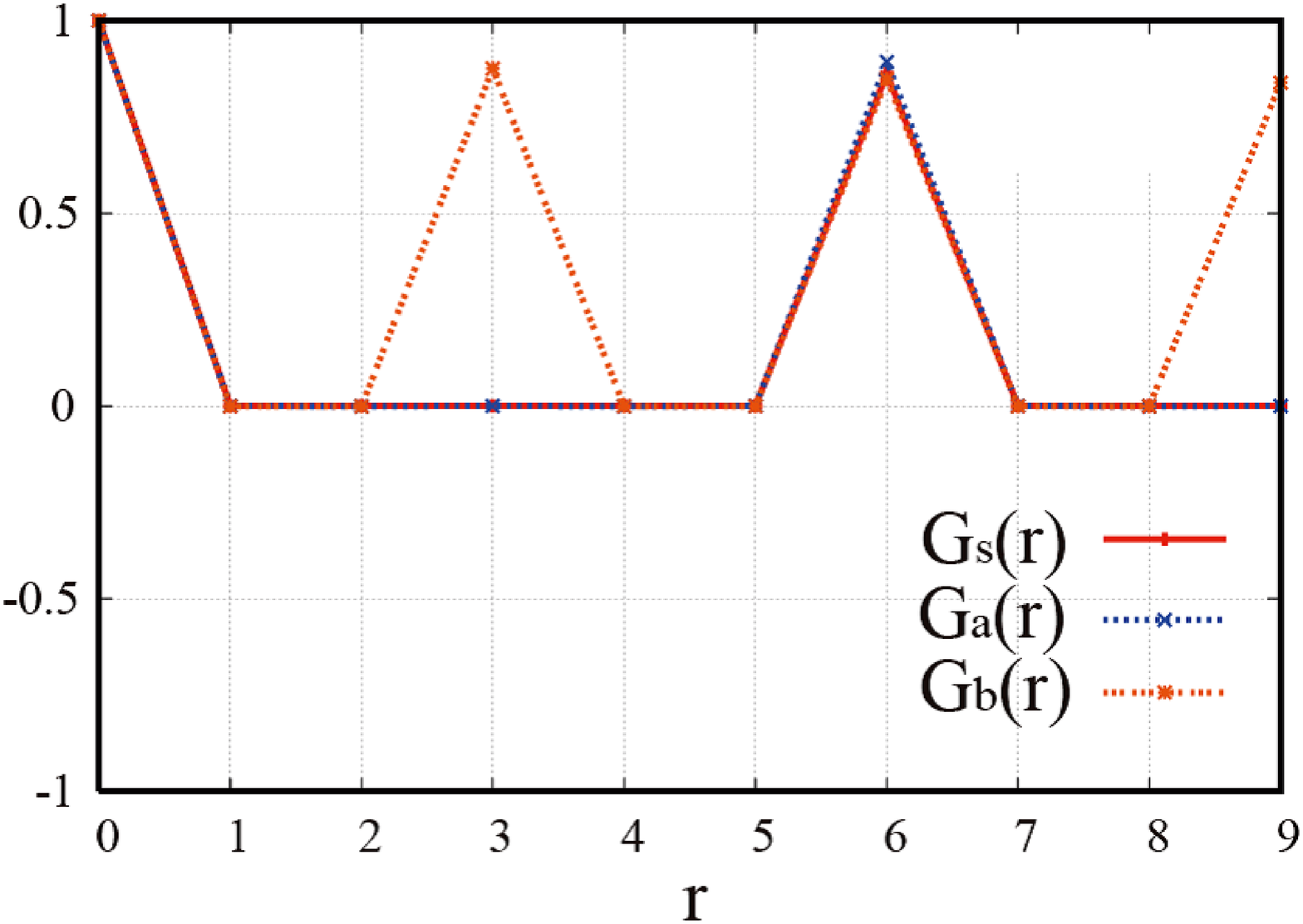}
\includegraphics[width=5cm]{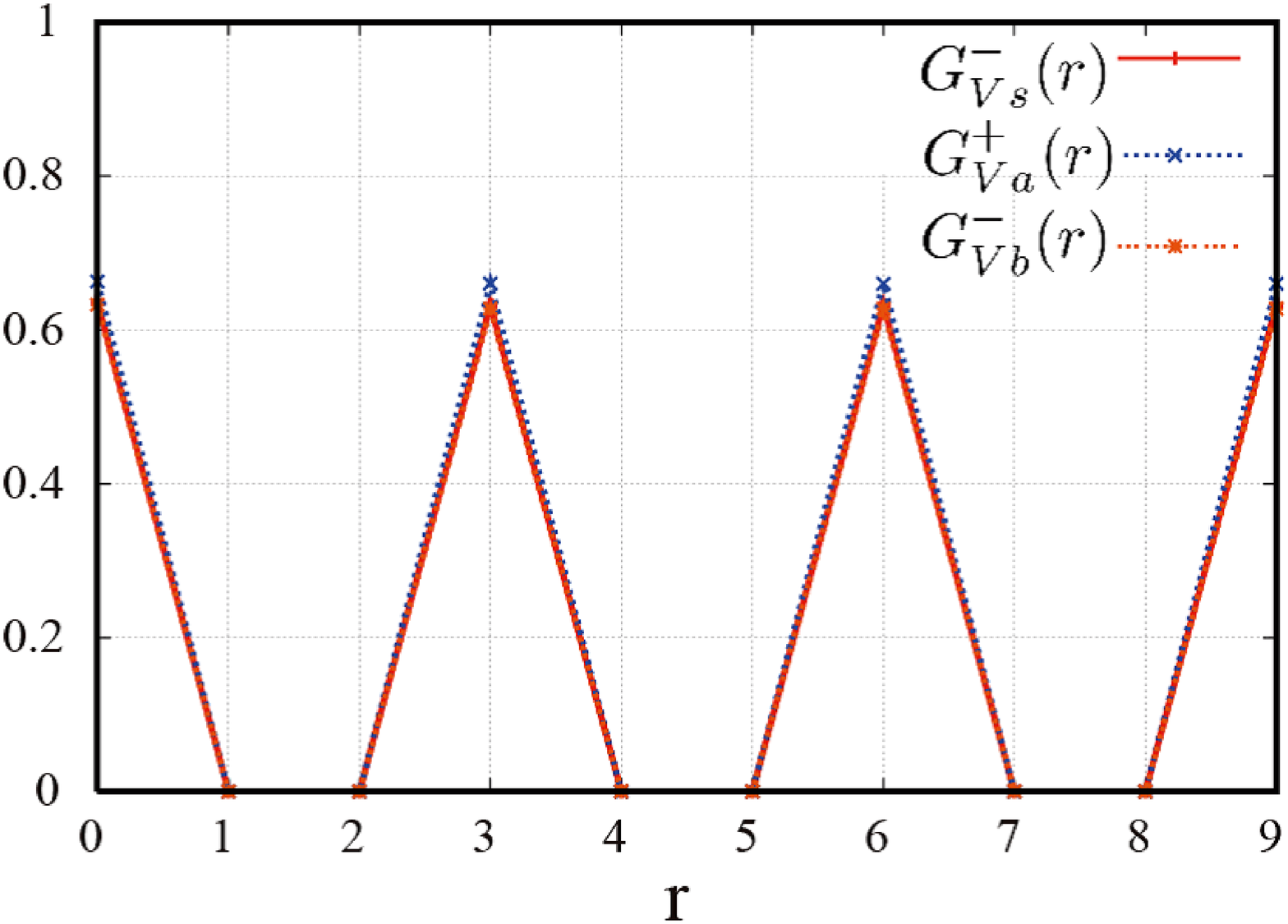}
\vspace{-0.3cm}
\caption{(Color online)
Correlation functions for the system with the mass difference at $f=1/3$.
$t_a=t_b/2$.
}
\label{correlations_md}
\end{center}
\end{figure}

To understand the behavior of the boson correlation functions,
the approach using the effective field theory in Sec.IV is useful.
The vector potentials couple with the field $\Phi_{\alpha i}$ 
though the hopping terms as in Eq.(\ref{AL2V}).
In the present case $f=f'$, the spin collective field $\Phi_{si}$
does not couple with the vector potential and then it can
have the LRO as the above numerical study indicates.
On the other hand, the fields $\Phi_{ai}$ and $\Phi_{bi}$couple to 
the vector potentials $A^a_{r,r'}$ and $A^b_{r,r'}(= A^a_{r,r'})$, respectively.
{\em One-body problem} in a constant magnetic field was studied by 
Hofstadter\cite{HBat}.
It was found that 
for general value of the magnetic field per unit plaquette,
fractal bands appear, whereas for rational number $f=p/q$ 
($p, q$ are integers and prime with each other) the energy spectrum
splits into $q$ bands and the ground state becomes $q$-fold degenerate.
This fact implies that the superfluid in the present case is a superposition
of the degenerate $q$ condensates whose phase degrees of freedom has
spatial dependence.
Interference between them causes cancellation of the correlation, and
the correlator has a nonvanishing value only in the case where the condensates
have the same phase.

Let us turn to the case with the mass difference $t_a=t_b/2$.
In this case, $f'=2f$ and $A^a_{r,r'}-A^b_{r,r'}=-A^a_{r,r'}$.
$E$ and $C$ exhibit similar behavior to those shown in Fig.\ref{ECmag}.
Stable state exists for $f=1/3, \ 1/2$ and $2/3$ as in the previous case.
Vortex density is shown in Fig.\ref{vortex_density_md}.
This result can be expected from the action $A_{\rm L}$ in Eq.(\ref{AL2V})
and the calculation in Fig.\ref{vor_density}.
For $f=1/3$, the correlation functions of the phase fields and the vortices are
shown in Fig.\ref{correlations_md}.
As the $b$-atom feels $f'=2/3$, the behavior of the correlation functions 
corresponding to it is easily understood from the results of $f=2/3$ with
the same mass.

\subsection{qXY model in a staggered magnetic field}

\begin{figure}[h]
\begin{center}
\includegraphics[width=4cm]{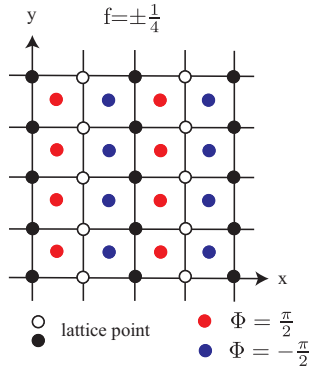}
\vspace{-0.3cm}
\caption{(Color online)
Square optical lattice with  staggered magnetic flux $\Phi=\pm \pi/2$
per plaquette.
}
\label{setup}
\end{center}
\end{figure}

In the previous subsection, we studied the qXY model in a uniform
magnetic field and found that the stable ground state forms for
specific strength of the magnetic field, and vortex lattice is realized there.
The vortex lattice is expected to be observed by the density profile of BEC.
In the experiments, two-dimensional 
optical lattice system in a strong staggered magnetic
field was realized and interesting phenomena were observed\cite{staggered}. 
Among them, spatial distribution of the ground-state phase
was observed for the staggered magnetic field with $\pm \pi/2$ per
plaquette. See Fig.\ref{setup}.
This system is closed related with the system studied in this paper,
in particular, in the ground-state properties.
The qXY model in a staggered magnetic field can be studied straightforwardly
as in the previous subsection.
The Bose-condensed ground-state is closely related with the ground-state
of the one-particle system in a staggered magnetic field investigated experimentally.

\begin{figure}[h]
\begin{center}
\includegraphics[width=5cm]{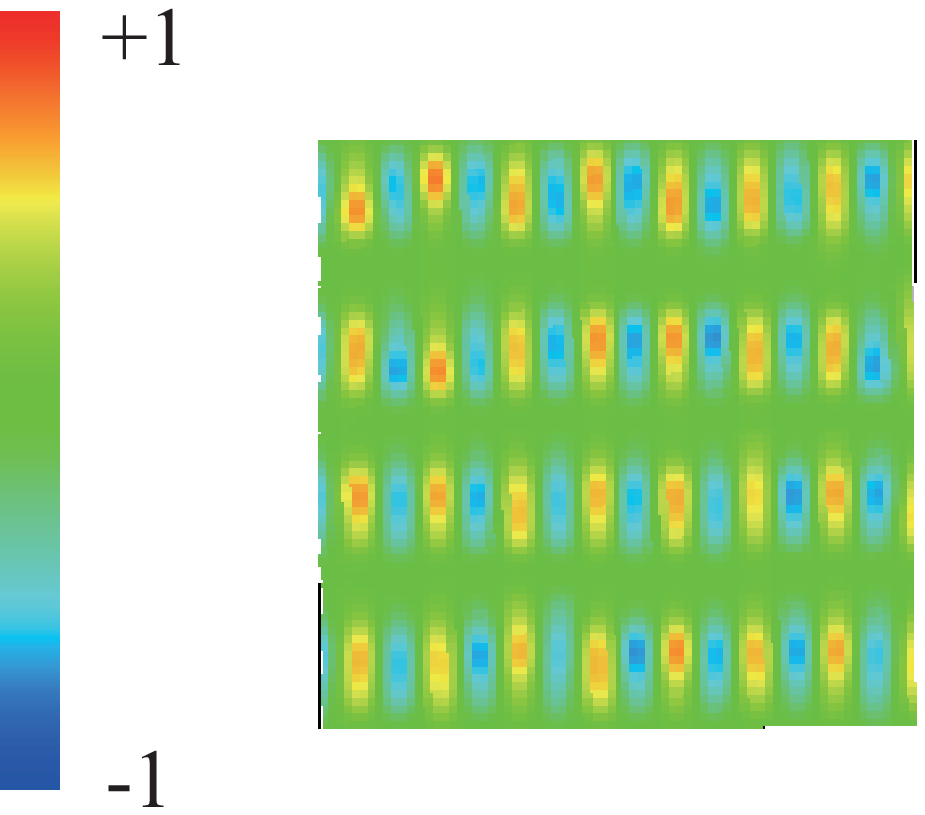}
\hspace{1cm}
\includegraphics[width=7cm]{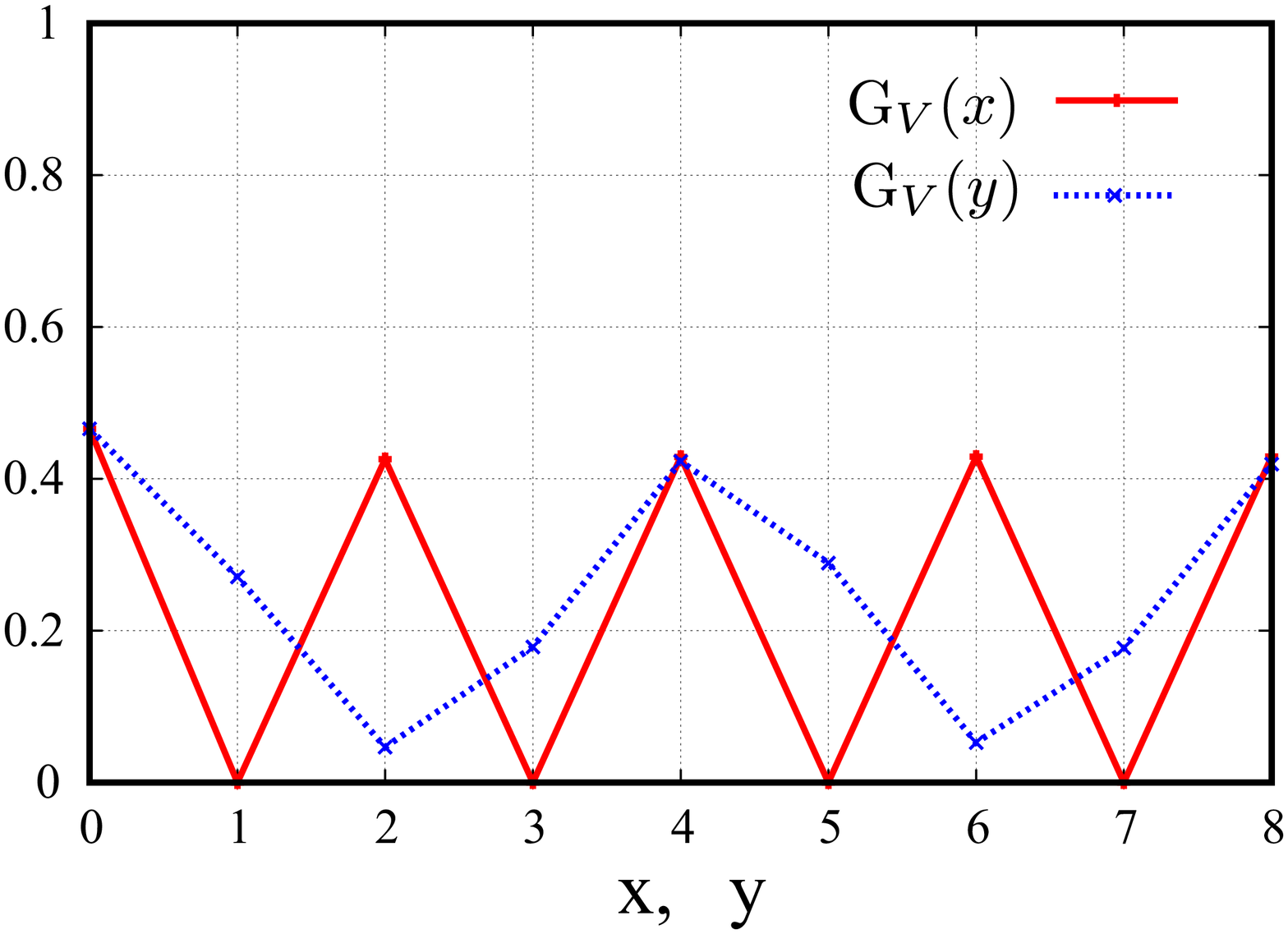}
\vspace{-0.3cm}
\caption{(Color online)
Snapshot and $+$vortex correlation functions for the system in $\pm \pi/2$
staggered magnetic field.
There is one vortex per four plaquettes along the $y$-direction.
The chirality of vortices alternates in the $x$-direction.
$-$vortex exhibits similar correlation.
}
\label{staggered}
\end{center}
\end{figure}

We focus on the case of single BEC system in the staggered magnetic
field $\pm \pi/2$ per plaquette and employ
the axial gauge for the vector potential, i.e. 
$A_{r,r+\hat{x}} \neq 0, A_{r,r+\hat{y}}=0$ as in the experimental setup
using the laser-assisted tunneling\cite{staggered}.
Furthermore to obtain the direct connection to the experimental
observation, we employ the definition of the vortex density Eq.(\ref{Vr})
instead of Eq.(\ref{VAr}).
In Fig.\ref{staggered}, we show snapshot of vortex lattice and also the vortex
correlation in both the $x$ and $y$-directions.
It is obvious that the vortex lattice forms and there is one vortex 
per four plaquettes along the $y$-direction.
The chirality of the vortices alternates in the $x$-direction due to
the staggering.
This result is essentially in good agreement with the experimental observation,
though the pattern of the vortex lattice is sightly different with each other.

In the previous subsection, we studied the BEC system in a uniform magnetic
field, and discussed how the obtained result is related with the Hofstadter butterfly.
We expect that a direct observation of the vortex lattice will be succeeded 
in near future.

\section{Conclusion}

In this paper, we studied the qXY model that describes dynamics of 
phase degrees of freedom of cold-atom fields in an optical lattice. 
The qXY mode is an effective low-energy model of the
bosonic t-J model and the Bose-Hubbard-J model.
By means of the MC simulations, we clarified the phase diagram of the qXY
model with a mass difference.
We found that there exist four phases and clarified critical behavior 
near the phase boundary.

We also considered the effects of the $J_z$-term in the bosonic t-J model,
in particular, we searched the parameter region of the SS in which both the spatial 
(checkerboard) and internal (SF) LROs coexist.
Then we derived the second form of the effective field theory by means of
the ``Hubbard-Stratonovich'' transformation, and we studied the NG bosons
and the Higgs mode and obtained interesting results.

Finally, we studied the qXY model in an external magnetic field.
We found that the BEC is easily destroyed by the external magnetic field,
but also at certain specific magnitudes of the magnetic field per plaquette, 
the stable SF states exist.
In these states, the vortices form nontrivial spatial lattice and the boson
correlation exhibits certain solid like order with a periodicity of a multiple 
lattice spacing.

As we explained in various places in the text, 
we hope that the above findings will be observed by experiments
of the cold atoms on optical lattices near future.


\acknowledgments 
This work was partially supported by Grant-in-Aid
for Scientific Research from Japan Society for the 
Promotion of Science under Grant No23540301.



\begin{thebibliography}{9}
\bibitem{cutoff}
In order to define the vortex density in Eq.(\ref{defV}), 
we have introduced a cutoff, and chosen its value to be $0.6$.
The cutoff is useful to identify the vortex at each site clearly.
\end{thebibliography}
\end{document}